\documentclass[12pt,preprint]{aastex}

\newcommand{\etal}{et al.}

\shorttitle{AGN in the Far IR}
\shortauthors{Draper \&  Ballantyne}

\begin{document}

\title{Properties and Expected Number Counts of Active Galactic Nuclei and their Hosts in the Far Infrared}


\author{A. R. Draper and D. R. Ballantyne}
\affil{Center for Relativistic Astrophysics, School of Physics,
  Georgia Institute of Technology, Atlanta, GA 30332}
\email{aden.draper@physics.gatech.edu}

\begin{abstract}
Telescopes like {\em Herschel} and the Atacama Large Millimeter/submillimeter Array (ALMA) are creating new opportunities to study sources in the far infrared (FIR), 
a wavelength region dominated by cold dust emission.  Probing cold dust in active galaxies allows for study 
of the star formation history of active galactic nuclei (AGN) hosts.  The FIR is also an important spectral region for observing 
AGN which are heavily enshrouded by dust, 
such as Compton thick AGN.  By using information from deep X-ray surveys and cosmic X-ray background synthesis 
models, we compute Cloudy photoionization simulations which are used to predict the spectral energy distribution 
(SED) of AGN in the FIR.  Expected differential number counts of AGN and their host galaxies are calculated in the 
{\em Herschel} bands. The expected contribution of AGN and their hosts to the cosmic infrared background (CIRB) 
and the infrared luminosity density are also computed.  Multiple star formation scenarios are investigated using a modified 
blackbody star formation SED.  It is found that FIR observations at $\sim$500 $\mu$m are an excellent tool in determining the star 
formation history of AGN hosts. Additionally, the AGN contribution to the CIRB can be used to determine whether 
star formation in AGN hosts evolves differently than in normal galaxies.  The contribution of Compton thick AGN to 
the bright end differential number counts and to the bright source infrared luminosity density is a good 
test of AGN evolution models where quasars are triggered by major mergers.

\end{abstract}

\keywords{galaxies: active --- galaxies: evolution --- galaxies: quasars: general --- infrared: galaxies --- X-rays: galaxies}

\section{Introduction}
\label{sect:intro}

Thanks to the {\em Spitzer Space Telescope} and its imaging and spectroscopic instrument suite \citep{Wer04}, it has recently been found that active galactic nuclei (AGN) can dominate the near and mid-infrared luminosity 
and make a significant contribution to the 8-1000 $\mu$m luminosity, $L_{IR}$, of AGN host galaxies 
\citep[and references therein]{H05, Y05, W06, SHW08}.  Specifically, a large fraction of ultra-luminous infrared galaxies 
($L_{IR}$ $>$ 10$^{12}$ L$_{\odot}$; ULIRGs) and luminous infrared galaxies ($L_{IR}$ $>$ 10$^{11}$ L$_{\odot}$; 
LIRGs) appear to host luminous AGN \citep[e.g.,][]{Dey08, D08, F08, F09, Y10}.  
In the far infrared (FIR) it is more difficult 
to determine the AGN contribution to the combined AGN and host observed flux as little 
is known about the AGN spectral energy distribution (SED) at these wavelengths.  For the purposes of this study we define the FIR as 70-1000 $\mu$m.

The next several years will see a huge growth in high quality FIR and submillimeter spectral data due to telescopes like the {\em Herschel Space Observatory} and the Atacama Large Millimeter/submillimeter Array (ALMA).  {\em Herschel}, launched in May 2009, offers unprecedented coverage of the FIR and submillimeter spectral regions \citep{Pil10}.  {\em Herschel} carries two imaging photometers, the Photodetector Array Camera and Spectrometer (PACS) which has photometric bands at 70, 100, and 160 $\mu$m \citep{Pog10} and the Spectral and Photometric Imaging Receiver (SPIRE) which has photometric bands at 250, 350, and 500 $\mu$m \citep{Gri10}.  As one of the primary science goals of {\em Herschel} is to study the evolution of galaxies \citep{Pil10}, three wide-area surveys of various depths are being undertaken.  The PACS evolutionary probe (PEP) will cover $\sim$3 deg$^{2}$ and will observe some fields, such as the GOODS fields, down to the depth of the 100 and 160 $\mu$m confusion limits of a few mJy \citep{B10}.  The {\em Herschel} Multi-tiered Extragalactic Survey (HerMES) will observe $\sim$70 deg$^2$ with SPIRE, reaching 5$\sigma$ depths of a couple mJy at 250 $\mu$m in some fields \citep{O10, Rose10}.  Observing $\sim$510 deg$^2$ with both PACS and SPIRE, the {\em Herschel} Atlas (H-ATLAS) survey is the largest open time survey being undertaken by {\em Herschel} \citep{Eales10}.  H-ATLAS will reach 5$\sigma$ depths of 67 mJy at 100 $\mu$m and 53 mJy at 500 $\mu$m \citep{Eales10, Ibar10, Pas10}.  The ground-based instrument Submillimeter Common-User Bolometer Array 2 (SCUBA-2) will operate at 450 and 850 $\mu$m and will offer an intermediate sensitivity and mapping speed between those of {\em Herschel} and ALMA \citep{Holland06}. Also ground-based, ALMA is an imaging and spectroscopic instrument operating in the millimeter and submillimeter regime.  The ALMA primer\footnote{available at http://www.almaobservatory.org/images/stories/publications} notes that at full science operations, starting in late 2012, ALMA will have a 450 $\mu$m band (Band 9) with a continuum sensitivity of 0.69 mJy and a 350 $\mu$m band (Band 10) with a sensitivity of 1.1 mJy, along with bands at longer wavelengths\footnote{Sensitivities are for an integration time of 60 seconds and a spectral resolution of 1.0 km/s.}.  The 450 $\mu$m band will be available for early science operations in mid 2011, albeit at reduced sensitivity.  {\em Herschel}'s observing capabilities, and in particular the SPIRE deep and wide surveys at 250, 350 and 500 $\mu$m, will provide promising candidate targets for ALMA early science operations.

It is expected that AGN and their hosts will be readily detected in the FIR by {\em Herschel}.  \citet{Hat10} found 
that a third of their AGN sample had {\em Herschel} 
5$\sigma$ detections at 250 $\mu$m with $f_{250\mu m}$ $>$ 12.8 mJy, where $f_{250\mu m}$ is the 250 
$\mu$m flux.  The FIR flux from AGN and their hosts will be due to dust heated by a combination 
of AGN radiation and star formation.  Observations suggest that only 
3--9$\%$ of submillimeter galaxies are dominated by AGN emission in the submillimeter \citep{Laird10}.  However, 
a large portion of galaxies, nearly 50$\%$ of the galaxies in the infrared selected sample of \citet{Y10}, show 
strong infrared emission from both star formation and AGN activity. Consequently, the AGN 
contribution to the FIR emission of AGN hosts must be considered in order to not over-estimate 
star formation rates in AGN hosts.

As the FIR SED of AGN is expected to generally be dominated by star formation \citep[e.g.][]{Hat10, L10}, 
FIR observations will inform investigations of the star formation history of AGN hosts.  By comparing 
the star formation history of AGN hosts and normal galaxies, the role of AGN in galaxy evolution will be constrained.
Understanding AGN host star formation history is therefore an important tool 
in determining what processes trigger AGN.  For example, \citet{S88} proposed that galaxy mergers trigger both 
intense starbursts and quasar activity.  Recent observations and simulations support this model 
\citep{P04, H06, Rig09, DB10, K10}.  \citet{P04} found that type 2 AGN tend to have higher 
SFRs than type 1 AGN.  \citet{DB10} found that a significant fraction of Compton 
thick (CT) AGN, AGN with an X-ray obscuring column density $N_H$ $>$ 10$^{24}$ cm$^{-2}$, are quasars which 
are accreting very rapidly, as predicted by simulations by \citet{H06} and others.  However, it appears 
that this model may only be applicable for the most powerful AGN and that Seyfert strength AGN may 
evolve more secularly \citep{B06a, Has08, L10, N10}.  In order to determine the applicability of the merger 
driven evolution and secular evolution models, the star formation history of AGN hosts must be 
accurately determined.  This requires a robust method of identifying AGN hosts and the AGN 
contribution to the FIR emission.

FIR emission from AGN comes from the "torus" of dusty gas, which, according to the unified scheme, surrounds the central engine of all AGN \citep[e.g.,][]{A93}.
This dusty torus will absorb higher energy emission from the central engine of the AGN and re-radiate this 
energy in the infrared.  The temperature of the dust, and therefore the peak wavelength of the infrared 
radiation, is dependent on the density of the obscuring gas and the distance of the obscuring gas from the 
central engine \citep[e.g.,][]{B06}, among other parameters like the geometry of the obscuring torus 
\citep[e.g.,][]{H09}.  It appears that the infrared SED of most X-ray selected AGN peaks in the mid-infrared 
\citep[e.g.,][]{E94, N07}.  However, due to the large amount of dust required to reach CT levels of obscuration, 
it is expected that the clouds of dusty gas in CT AGN will have a greater spatial extent than in less obscured AGN.  
Therefore, CT AGN will have a significant reservoir of dusty gas which is cooler than most AGN-related dust 
clouds.  This cooler dust component will cause CT AGN to be brighter in the FIR than unobscured AGN.

Indeed, studies suggest that bright galaxies detected in the FIR host a large number of heavily obscured, 
$N_H$ $>$ 10$^{23}$ cm$^{-2}$, and possibly CT AGN \citep[e.g.,][]{A05a, A05b, Bon10, W10}.  
\citet{Mul10} found that 45--75$\%$ of the 1 Ms CDF-S X-ray sources will be 
detected by {\em Herschel} at 100 $\mu$m and that deep infrared observations with {\em Herschel} will allow for 
a significant fraction of the CT AGN population to be identified.  Also, since X-ray selection misses 
nearly half of all infrared identified AGN \citep{Fu10} it is expected that FIR telescopes like {\em Herschel} 
and ALMA will detect highly obscured AGN that are missed by the deep X-ray surveys.  However, there is an open debate 
as to how many of these infrared AGN reach CT levels of obscuration.  \citet{F09} find that up to 90$\%$ of 
sources with $f_{24}$ $>$ 550 $\mu$Jy and $f_{24}/f_R$ $>$ 1000 are CT AGN, where $f_{24}$ is the 24 $\mu$m flux and $f_R$ 
is the R-band flux.  Conversely, \citet{Georg10} find a more moderate fraction of CT AGN for this 
population.  When looking at similar infrared excess sources, \citet{Gk10} found no strong evidence suggesting 
that these sources host AGN with CT levels of obscuration.  Therefore further study of the infrared properties 
of AGN are important in determining exactly how many CT AGN are hidden within dusty luminous infrared galaxies.  

This work makes predictions for FIR AGN number counts, contribution to the CIRB, and luminosity density using a 
population synthesis model informed by constraints from the cosmic X-ray background (CXRB) and deep X-ray surveys, including the Eddington 
ratio dependent CT fraction, $f_{CT}$, of the composite model investigated by \citet{DB10}.  The effect of 
various AGN host star formation scenarios are also investigated.  The calculation of the model SEDs 
is discussed in \S \ref{sect:seds}.  Predictions for bare AGN are presented in \S 
\ref{sect:bare} followed by predictions for various host star formation scenarios in \S \ref{sect:sf}.  In 
\S \ref{sect:disc} and \S \ref{sect:sum} the results are discussed and summarized.  A $\Lambda$CDM 
cosmology is assumed as necessary, with $h_0$ = 0.7, $\Omega_m$ = 0.3, and $\Omega_{\Lambda}$ = 0.7 \citep{Hi09}.

\section{Calculation of AGN SEDs}
\label{sect:seds}

As the goal of this study is to make predictions of the average FIR properties of AGN, SEDs are used 
which are representative of an ensemble of AGNs at a given 2--10 keV luminosity, $L_{X}$, and $z$, instead of using a SED 
template based on observations of a statistically small set of AGNs.  Since AGN IR emission is primarily 
due to the obscuring gas and dust re-radiating absorbed X-ray emission, photoionization simulations allow 
for the computation of the IR emission of an average AGN with a given $L_{X}$ 
and $N_{H}$.  Also, this SED computation method provides an opportunity 
to explore the parameter space of obscuring gas location, density distribution, and dust 
content, which, upon comparison with observations, may offer constraints for these physical parameters.

Similarly to \citet{B06}, the IR AGN SEDs are calculated using the photoionization code Cloudy version C08.00 \citep{F98}. 
Cloudy includes the complicated physics of radiative transfer through dusty gas and uses a physical dust 
model which takes into account silicate and graphite grains along with polycyclic hydrocarbons (PAHs).  This 
technique does necessitate a simplification of the IR emitting region, which might actually be quite complex 
\citep[e.g.,][]{H09,H10}.  However, since our purpose 
is to describe average properties of AGN, and not to model individual objects, this technique is an appropriate 
method of SED calculation. This assumption will be tested by comparing the model SEDs against real data in 
\S \ref{sub:prop}.

\subsection{Cloudy Model Setup}
\label{sub:setup}

Each Cloudy model is setup such that a constant AGN spectrum, characterized by $L_{X}$, 
is incident upon a cloud with inner radius $r$ from the continuum source.  The inner radius 
$r$ is set to 10 pc.  Compton thin clouds are assigned a uniform hydrogen density $n_H$ = 
10$^4$ cm$^{-3}$, as is typical of molecular clouds.  In order to prevent the dust mass 
from becoming too large, the Compton thick clouds are assigned $n_H$ = 10$^6$ cm$^{-3}$.  
Molecular clouds of this density are not uncommon and have been observed in the Large 
Magellanic Cloud \citep{R09} and in the Orion Nebula \citep{P07}.
Simulations were also run with $r$ = 1 pc and $n_{H}$ = 10$^4$ cm$^{-3}$ for both 
Compton thin and CT clouds to test the sensitivity of the results to these 
assumptions, which is discussed later.  

The Cloudy model input files are similar to those used by \citet{B06}, 
with the following improvements. Instead of using a constant $\alpha_{OX}$ for all AGN, the \citet{S06} 
$\alpha_{OX}$-L$_{2keV}$ relation is used here to determine $\alpha_{OX}$.  The most significant improvement made includes 
dividing the AGN population into Eddington ratio bins, high ($L/L_{Edd}>0.9$, where $L$ is the bolometric luminosity found using 
the bolometric correction by \citet{Mar04} and $L_{Edd}$ is the Eddington luminosity), moderate ($0.9<L/L_{Edd}<0.01$), 
and low ($0.1<L/L_{Edd}$), as described by \citet{DB10}.  Both the composite and original models of \citet{DB10} 
are investigated.  The $f_{CT}$ of the composite model is Eddington ratio dependent and finds that $\sim$86$\%$ of AGN 
accreting at greater than 90$\%$ of their Eddington rate and $\sim$60$\%$ of AGN accreting at less than 1$\%$ 
of their Eddington rate are Compton thick.  The original model assumes Compton thick AGN are a simple extension 
of the Compton thin type 2 population and that $\sim$44$\%$ of type 2 AGN are Compton thick.  Since the covering 
factor is set assuming the unified scheme holds, the covering factors are Eddington ratio dependent since 
$f_{CT}$ is Eddington ratio dependent.  For Compton thick objects the covering factor is set as the Compton thick fraction, $f_{CT}$.    
For objects with $22 \leq \log N_H < 24$ the covering factor is set as $(1.0-f_{CT})f_2$ where $f_2$ is the type 2 
fraction and is calculated as discussed in \S 2.2 of \citet{DB09}.  The covering factor for objects with 
$20 \leq \log N_H < 22$ is set to $(1.0-f_{CT})(1.0-f_2)$.  As the covering factor is dependent on the Eddington 
ratio and varies from $z$ = 0 to 1, Cloudy models had to be 
calculated as a function of $z$ and $L/L_{Edd}$ for each $L_X$ and $N_H$. 

There is evidence that different levels of obscuration in quasars 
might be related by an evolutionary scenario instead of by orientation effects \citep[e.g.,][]{S88, P04, DB10, D10}, 
and thus the covering factor need not be related to $f_{CT}$ or $f_2$.  \citet{B06} found that the unified model 
assumption holds for lower luminosity quasars and Seyfert galaxies but does not seem to hold for high-luminosity quasars, 
possibly due to a different evolution for high luminosity, and therefore high Eddington ratio, quasars.  Here this issue 
is addressed by using the Eddington ratio dependent $f_{CT}$ of the composite model, which takes into account the different 
evolution of high Eddington ratio quasars from the moderate and low Eddington ratio AGN.

\subsection{The Model Grids}
\label{sub:grid}

As this investigation of the FIR properties of AGN is informed by the constraints offered by the CXRB and X-ray 
observations of AGN, SEDs are calculated as a function of $\log L_X$ and $z$.  It is assumed that the redshift 
evolution of $f_2$ halts at $z=1$, thus models are computed up to $z$ = 1 and the $z$ = 1 models are used at $z>$ 1.  
Therefore, we compute Cloudy models for $z$ = 0 to 1, in steps of 0.05, and $\log L_X$ = 41.5 to 48, in steps of 0.25. 
For each luminosity and redshift, Cloudy models are calculated for each $\log N_H$ = 20.0, 20.5, $\ldots$, 24.5, 
25.0 cm$^{-2}$.  Since the covering factor is Eddington ratio dependent, Cloudy models had to be calculated for 
each luminosity, redshift, and column density for each of the three Eddington ratio bins.  This resulted in 18711 
individual Cloudy models for each $r$ investigated and an additional 5103 Cloudy models to investigate the 
sensitivity of results to the hydrogen density, $n_H$, of the Compton thick clouds.  

As we need to run 33 models for each $(L_X,z)$ pair, it is computationally prohibitive to run models on a finer grid.  
Therefore, we have linearly interpolated between the SEDs to allow for a finer grid in the calculations.  Thus eight 
SEDs are interpolated between consecutive $\log L_X$ model SEDs at a given $z$ and two SEDs are interpolated between 
consecutive $z$ model SEDs at a given $L_X$.  A convergence test was conducted using twice 
as many steps in both $L_X$ and $z$; it was found that the step size used here is adequate for convergence.

\begin{figure*}
\begin{center}
\includegraphics[angle=0,width=0.95\textwidth]{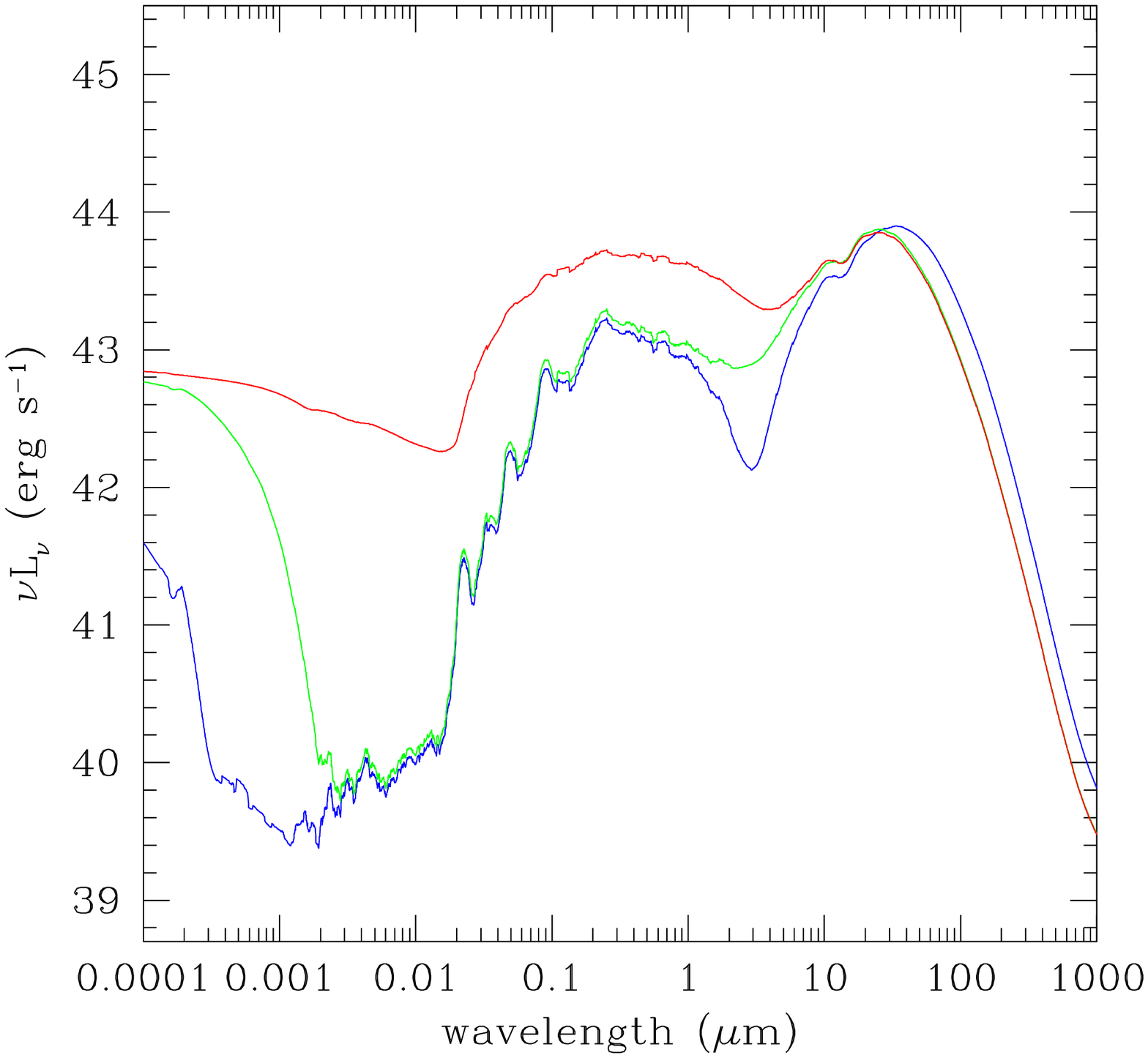}
\end{center}
\caption{Rest frame SEDs for high Eddington ratio AGN ($L/L_{Edd}$ $>$ 0.9) with $\log L_X$ = 43 and $z$ = 0.45 ($f_{CT}$ = 0.86 and $f_2$ = 0.78).  The type 1 SED is shown in red, the type 2 SED is shown in green, while the Compton thick SED is shown in blue.}
\label{fig:sed}
\end{figure*}

To create final SEDs from the 18711 Cloudy models, we follow the method described by \citet{B06}.  
For each Eddington ratio bin, the weighted average of the reflection components of each $N_H$ is 
added to the net transmitted continua of each $N_H$ to create the "unified SEDs".  The final SEDs 
are calculated in three categories: type 1, type 2 but Compton thin (which will be referred to 
as "type 2"), and Compton thick, for each Eddington ratio bin.  The type 1 SED is an average of the 
$20 \leq \log N_H < 22$ unified SEDs, the type 2 SED is an average of the $22 \leq \log N_H < 24$ 
unified SEDs, and the CT SED is an average of the $24 \leq \log N_H$ unified SEDs.  At this point, 
for each ($L_X$,$z$) pair there is a type 1, type 2, and Compton thick SED for each Eddington ratio bin.  
Figure \ref{fig:sed} shows the rest frame $\log L_X$ = 43 and $z$ = 0.45 SEDs for the high Eddington ratio 
bin, which has $f_{CT}$ = 0.86 and $f_2$ = 0.78.

\subsection{Properties of AGN Model SEDs}
\label{sub:prop}
The model SEDs were tested in the same manner as by \citet{B06} to assure the assumptions used to compute the 
SEDs were appropriate and that the model SEDs are consistent with observed SED trends.  As little is known 
about AGN FIR SEDs, these tests primarily investigate the mid-infrared properties of the model SEDs.  The 
mid-infrared colors were examined and found to be consistent with the findings of \citet{Brand06}.  The mid-infrared 
to X-ray flux ratio was considered as a function of $N_H$ and no correlation was found, in agreement with 
observational data \citep{L04, R04, G09}.  Also, the fluxes were tested against and found to be consistent 
with expectations from local bright AGN, known as "Piccinotti AGNs" \citep[see][]{AH04}.  As our 
goal is to make predictions which characterize an ensemble of AGNs, like number counts, the model SEDs are 
not compared against individual objects but against data which explores average properties of AGN SEDs.  The 
model SEDs are found to be consistent with the observed AGN SED trends.

\begin{figure*}
\begin{center}
\includegraphics[angle=0,width=0.95\textwidth]{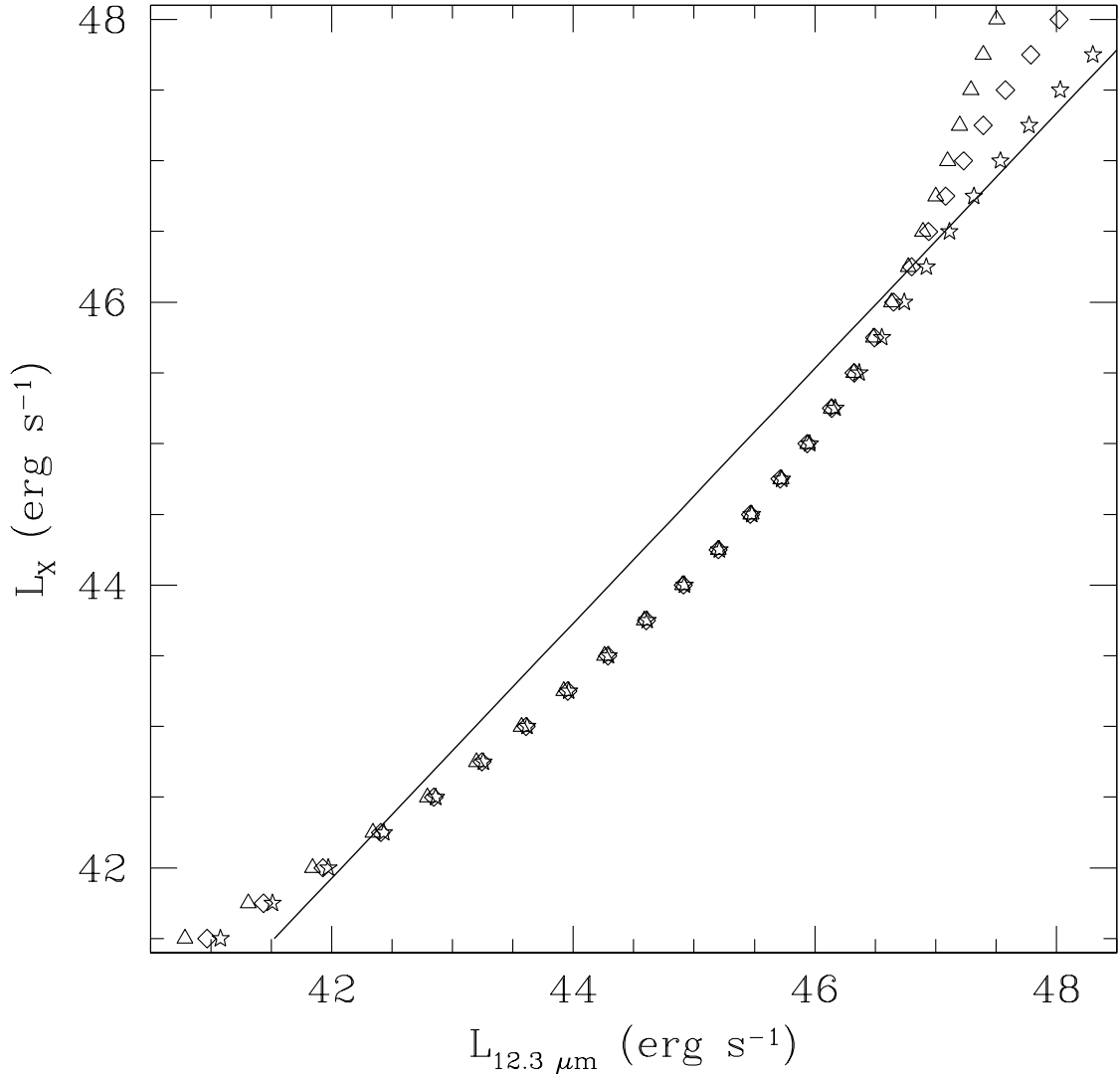}
\end{center}
\caption{$L_X$ versus $L_{12.3 \mu m}$ for the high Eddington ratio ($L/L_{Edd}$ $>$ 0.9) model SEDs at $z$ = 0.05.  Type 1 SEDs are shown as stars, type 2 SEDs are shown as diamonds, and the Compton thick SEDs are shown as triangles.  The black line is the best-fit line for the well-resolved sample of \citet{G09}.}
\label{fig:mir}
\end{figure*}

The model SEDs were also tested against the more recently discovered correlation between mid-infrared luminosity, $L_{12.3 \mu m}$, 
and $L_X$ for a sample of local Seyferts \citep{G09}.  Figure \ref{fig:mir} compares the $z$ = 0.05 model 
SEDs to the local best-fit correlation found by \citet{G09}.  The model SEDs agree reasonably well with 
the correlation, especially the type 1 SEDs.  At high luminosities the Compton thick SEDs appear 
to have a slight X-ray excess compared to the local Seyfert correlation.  This is due to the 
Compton thick SEDs including more power radiated in the FIR, and consequently less power radiated in the 
mid-infrared, than the Compton thin SEDs.  The $r$ = 1.0 pc model SEDs do match the $L_{12.3\mu m}$ -- $L_X$ 
relation better than the $r$ = 10 pc model SEDs; however, the $r$ = 10 pc model SEDs best describe the 
general mid-infrared properties of an ensemble of AGN \citep[see][]{B06}.  The fact that the $r$ = 1.0 pc model 
SEDs fit this relation better than the $r$ = 10 pc model SEDs shows that the distribution of gas and dust in the 
AGN dusty torus is an important effect in the mid-infrared.  In order for the models used here to better approximate 
the \citet{G09} relation, the geometry of the dusty torus should be carefully taken into account.  The FIR SEDs are dominated by star formation, so the geometric details of the hot dust close to the central engine should not affect the results of this study.  Also, as this study is 
focused on the integral properties of a large number of AGN, detailed modeling of the torus geometry is beyond 
the scope of this current study and we leave this for future work.   Overall, the model SEDs are an appropriate representation 
of average AGN SEDs.

\section{Predictions for Bare AGNs}
\label{sect:bare}
In this section we present predictions for differential number counts, the AGN contribution to 
the cosmic infrared background (CIRB), and luminosity density, based on the model SEDs.  The population 
synthesis model last described by \citet{DB10} is used to incorporate the information about and 
the constraints on the AGN population from deep hard X-ray surveys and the CXRB into infrared predictions. 

\subsection{Differential Number Counts}
\label{sub:counts}
The number of sources per square degree with flux greater than $S$, $N(>S)$ is found by
\begin{equation}
N(>S) = \frac{K^{deg}_{sr}c}{H_0} 
\times \int^{z_{max}}_{z_{min}} \int^{\log L_X^{max}}_{max(\log L_X^{min}, \log L_X^S)} \frac{d\Phi_{\lambda} (L_X, z)}{d\log L_X} \frac{d_l^2}{(1+z)^2[\Omega_m(1+z)^3+\Omega_{\Lambda}]^{1/2}} d\log L_X dz,
\label{eq:counts}
\end{equation}
where the factor $K^{deg}_{sr}$ = 3.05 $\times$ 10$^{-4}$ converts from sr$^{-1}$ to deg$^{-2}$, 
$d\Phi_{\lambda}(L_X,z)/d\log L_X$ is the evolving Eddington ratio space density computed by \citet{DB10}, 
in Mpc$^{-3}$, $d_l$ is the luminosity distance, and $\log L_X^S$ is the 2--10 keV rest-frame luminosity 
corresponding to the observed-frame infrared flux $S$ at redshift $z$.  The differential number counts are 
found by taking the derivative of $N(>S)$ with respect to $S$, $dN(>S)/dS$.  The differential number 
counts were also calculated directly.  However, due to the coarseness of the $L_X$ and $z$ grid used, the 
differential number counts contained numerical artifacts at brighter fluxes.  The predictions from the two calculation methods 
are in agreement; however, computing the integral number counts on a very fine flux grid and then 
taking the derivative with respect to flux minimizes the numerical noise found in direct 
calculation of the differential number counts.

\begin{figure*}
\begin{center}
\includegraphics[angle=0, width=0.95\textwidth]{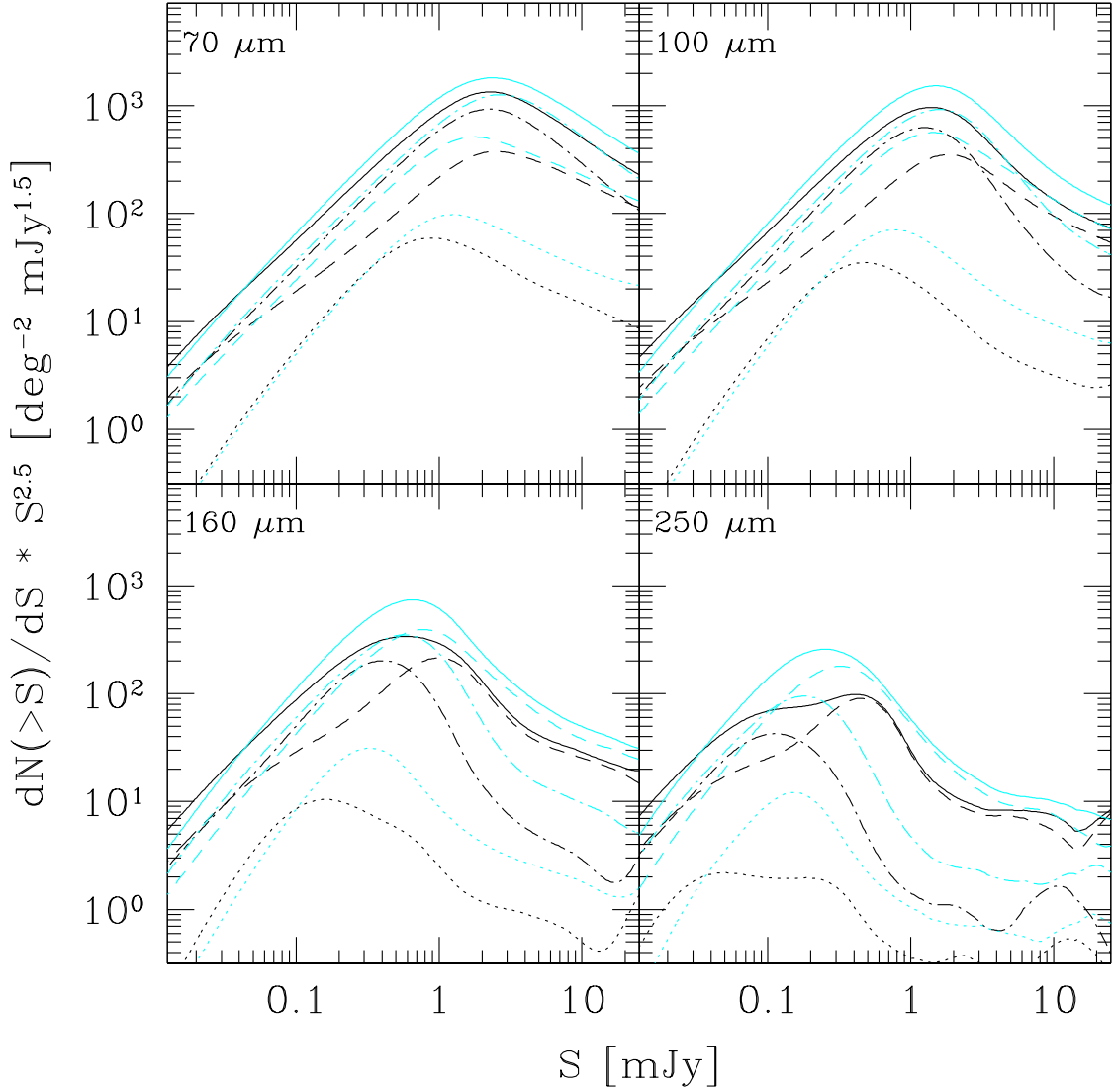}
\end{center}
\caption{Euclidean normalized differential number counts for bare AGN at 70, 100, 160, and 250 $\mu$m.  The black lines plot the predictions based on the composite model of \citet{DB10} and the cyan lines show the predictions for the original model.  The dotted-lines are predictions for type 1 AGN, the dot-dashed lines are predictions for type 2 AGNs, and the dashed lines are predictions for Compton thick AGN.}
\label{fig:baredNdS}
\end{figure*}

The Euclidean normalized differential number counts for 70, 100, 160, and 250 $\mu$m are shown in 
figure \ref{fig:baredNdS}\footnote{Fluxes are calculated at the filter nominal wavelengths.  Using the full filter transmission function to calculate the dN/dS provides a result which is within a factor of 1.05 of that obtained when calculating the flux at the filter nominal wavelength.  Thus the error due to calculating the fluxes at the filter nominal wavelength instead of using the full filter profile is negligible compared to the uncertainties in the model and the measurements.}.  Plots are not shown for the 350 and 500 $\mu$m 
bands because, as evident in figure \ref{fig:sed}, the flux due to the AGN is very small at such long wavelengths; for example, at 10 mJy the 500 $\mu$m band 
$dN(>S)/dS$ $\approx$ 0.4 AGN hosts mJy$^{1.5}$ deg$^{-2}$.  The black lines show the predictions for the composite model and the 
cyan lines show the predictions for the original model.  For ease in interpreting the figure, the 
low, moderate and high Eddington ratio bins have been combined and the total differential counts 
are shown (see \S \ref{sub:edd} for a discussion about the contribution from the various Eddington ratio bins).  
As expected, the differential number counts for bare AGN are very small at long wavelengths because the 
AGNs tend to create hot dust.  Therefore, long wavelengths can be used to investigate the star formation 
in AGN hosts.  For both the original and composite models, CT AGN dominate the bright end of the 
differential counts, especially at wavelengths longer than 70 $\mu$m. 
In both models the type 2 AGN dominate at lower fluxes and the type 1 AGN contribute significantly less 
than their obscured counterparts at all wavelengths. The original model and composite model differential 
counts peak at about the same flux for each wavelength and have similar bright end slopes.  However, 
the original model peaks above the composite model and declines faster on the lower flux level end.

\begin{figure*}
\begin{center}
\includegraphics[angle=0, width=0.95\textwidth]{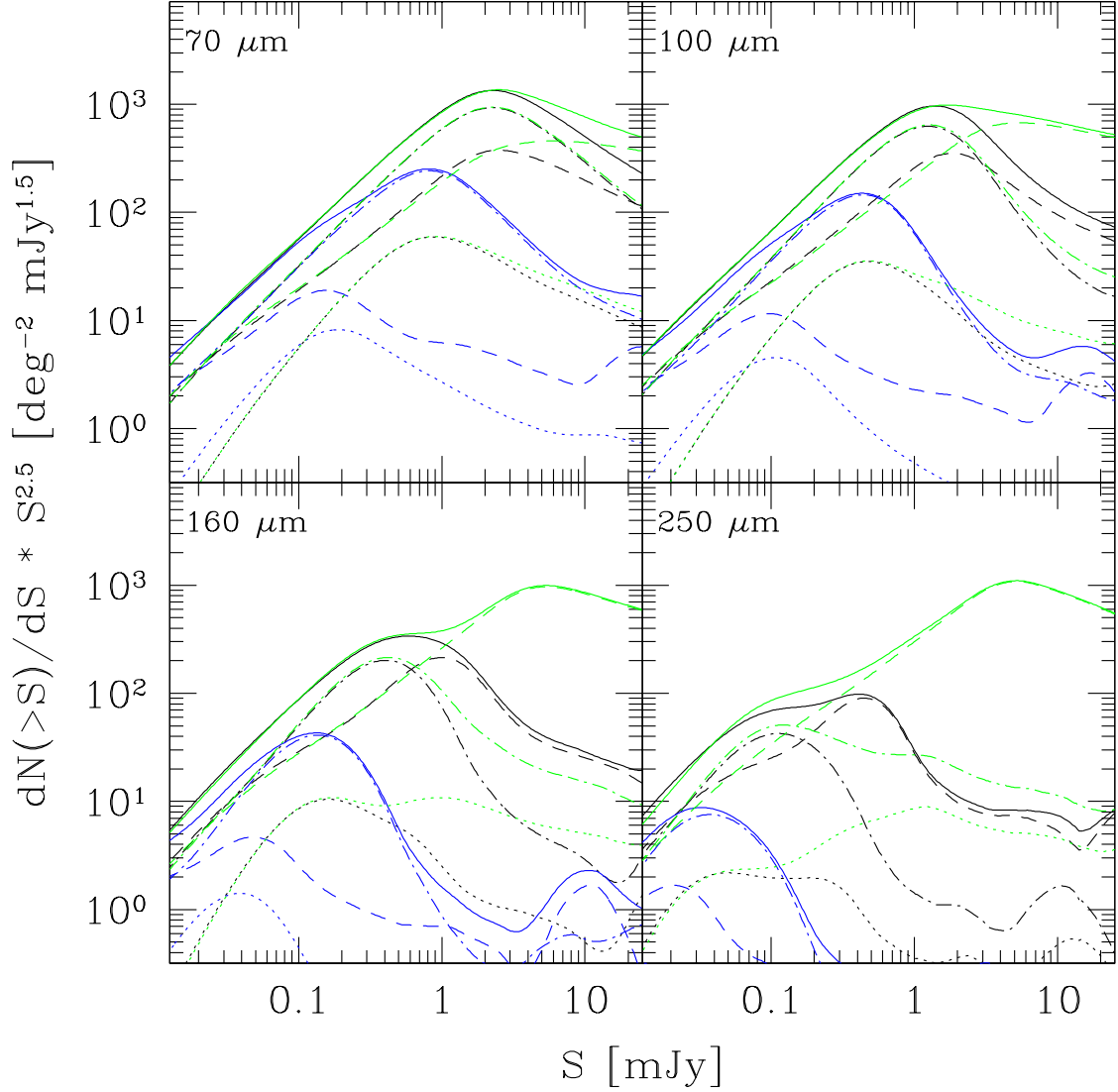}
\end{center}
\caption{Comparison of Euclidean normalized differential number counts for bare AGN at 70, 100, 160, and 250 $\mu$m using the composite model.  The blue lines are $r$ = 1 pc and $n_{H, CT}$ = 10$^{6}$ cm$^{-3}$, the black lines are $r$ = 10 pc and $n_{H, CT}$ = 10$^{6}$ cm$^{-3}$, and the green lines are $r$= 10 pc and $n_{H, CT}$ = 10$^{4}$ cm$^{-3}$.  The line styles are the same as in figure \ref{fig:baredNdS}.}
\label{fig:baredNdSr}
\end{figure*}

It is important to consider the dependence of these predictions on the parameters of the distance of 
the inner edge of the cloud from the ionizing source, $r$, and the  $n_H$ of the CT clouds, $n_{H,  CT}$.
As seen in figure \ref{fig:baredNdSr}, when $r$ = 1 pc the differential counts are greatly reduced 
at all FIR wavelengths because the dust is hot and radiating more energy in the mid-infrared.  Contrastingly, 
when $r$ $>$ 10 pc the differential counts are enhanced due to the dust being farther from the illuminating 
source and therefore cooler.  Similarly, when $n_{H, CT}$ = 10$^4$ cm$^{-3}$ 
the differential counts are slightly enhanced in the FIR due to the greater spatial extent of the cloud.  
However, since the combined AGN and host SED will be dominated by star formation at longer wavelengths, 
and there is likely to be variation among individual objects, the exact values for $r$ and $n_{H,CT}$ used 
will have little effect on the final predictions.

\subsection{CIRB}
\label{sub:cirb}
\begin{figure*}
\begin{center}
\includegraphics[angle=0, width=0.95\textwidth]{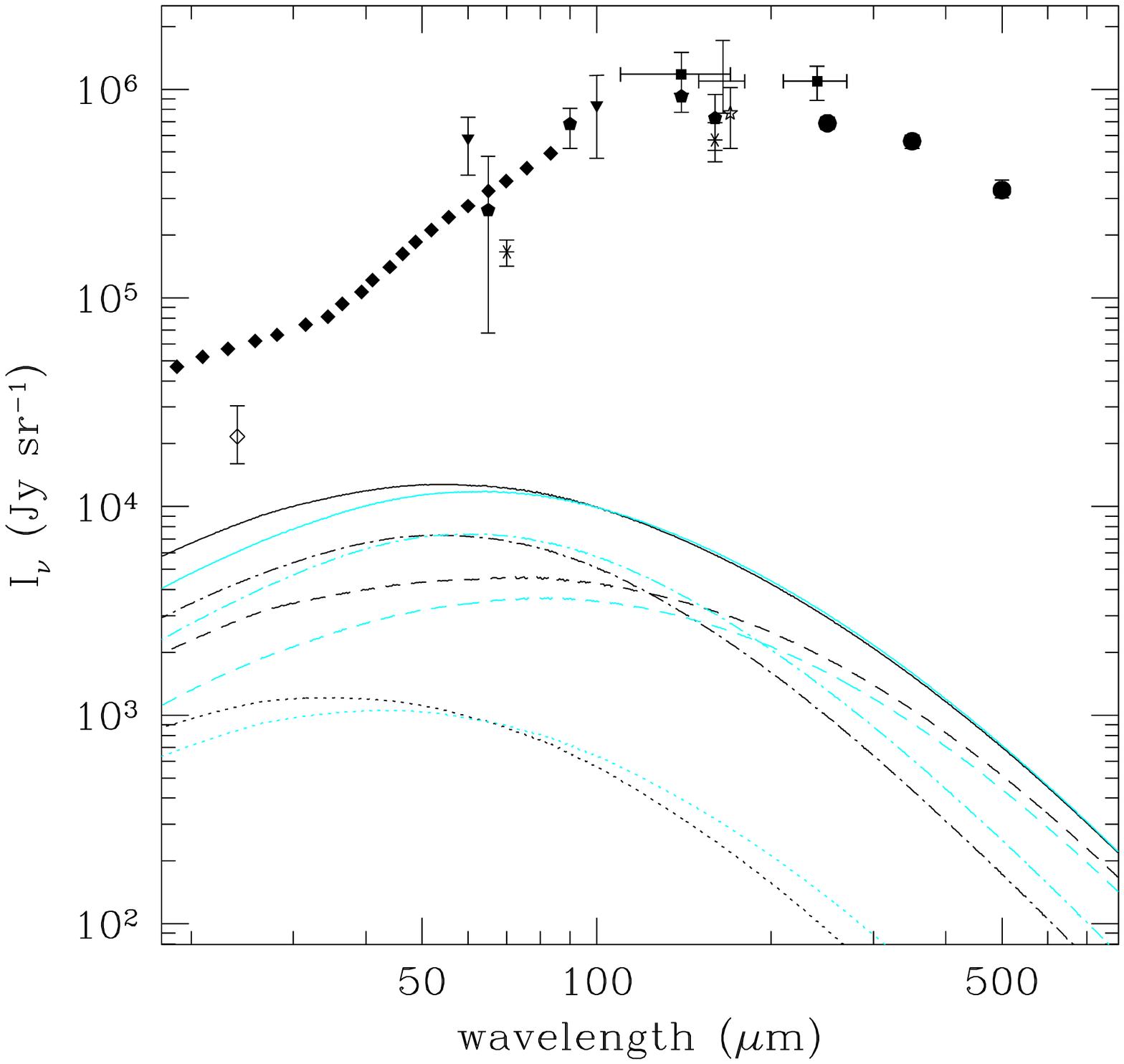}
\end{center}
\caption{Bare AGN contribution to the CIRB.  Line color and styles are the same as in figure \ref{fig:baredNdS}.  Data points are from a variety of instruments: the filled diamonds are from IACTS \citep{MR07}; the open diamond is from MIPS \citep{Pap04}; the triangles are from DIRBE \citep{F00}; the pentagons are from AKARI \citep{M10}; the asterisks are from MIPS \citep{D06}; the squares are from DIRBE \citep{H98}; the cross is from ISOPHOT \citep{J09}; the star is also from ISOPHOT \citep{LP00}; and the circles are from BLAST \citep{P09}.}
\label{fig:cirb}
\end{figure*}

The AGN contribution to the CIRB is calculated similarly to population synthesis models looking at the 
CXRB, but exchanging the X-ray spectrum with the infrared spectrum.  Thus we have
\begin{equation}
I_{\nu}(\nu)=\frac{c}{H_0} \int_{z_{min}}^{z_{max}} \int_{\log L^{min}_{X}}^{\log L^{max}_X} \frac{d\Phi_{\lambda} (L_X,z)}{d\log L_X} \times \frac{S_{\nu}(L_X,z) d_l^2}{(1+z)^2[\Omega_m(1+z)^3+\Omega_{\Lambda}]^{1/2}} d\log L_X dz,
\label{eq:cirb}
\end{equation}
where $S_{\nu}(L_X,z)$ is the observed-frame AGN model SED, in Jy sr$^{-1}$, computed with $L_X$ at redshift $z$.
As there is little knowledge about the SEDs of AGN in the FIR, there are very few constraints on the 
AGN contribution to the CIRB.  \citet{J10} claim that AGN contribute $\lesssim$ 10$\%$ of the CIRB 
at $z$ $<$ 1.5, based on {\em Spitzer} observations in the GOODS and COSMOS fields.  By extrapolating from 
CXRB models to the CIRB using rough AGN SED templates, \citet{S04} predict that bare AGN contribute 
$\sim$0.3$\%$ of the CIRB at 160 $\mu$m\footnote{Here we assume the intensity of the CIRB at 160 $\mu$m 
is 12.84 nW m$^{-2}$ sr$^{-1}$ \citep{A10}.}.  Using a method similar to the one described here, 
\citet{BP07} predict that bare AGN contribute $\sim$0.9$\%$ of the CIRB at 160 $\mu$m.  Here it is found 
that both the composite and original models predict that bare AGN contribute $\sim$0.9$\%$ of the CIRB at 
160 $\mu$m.  As seen in figure \ref{fig:cirb}, type 2 AGN dominate the CIRB at lower wavelengths and at 
higher wavelengths CT AGN dominate the AGN contribution to the CIRB.  At wavelengths greater than $\sim$100 
$\mu$m the original model and the composite model predict very similar contributions of AGN to the CIRB.  
Below $\sim$100 $\mu$m the composite model predicts a higher contribution to the CIRB by AGNs than the 
original model.

\subsection{Luminosity Density}
\label{sub:lum}
\begin{figure*}
\begin{center}
\includegraphics[angle=0, width=0.95\textwidth]{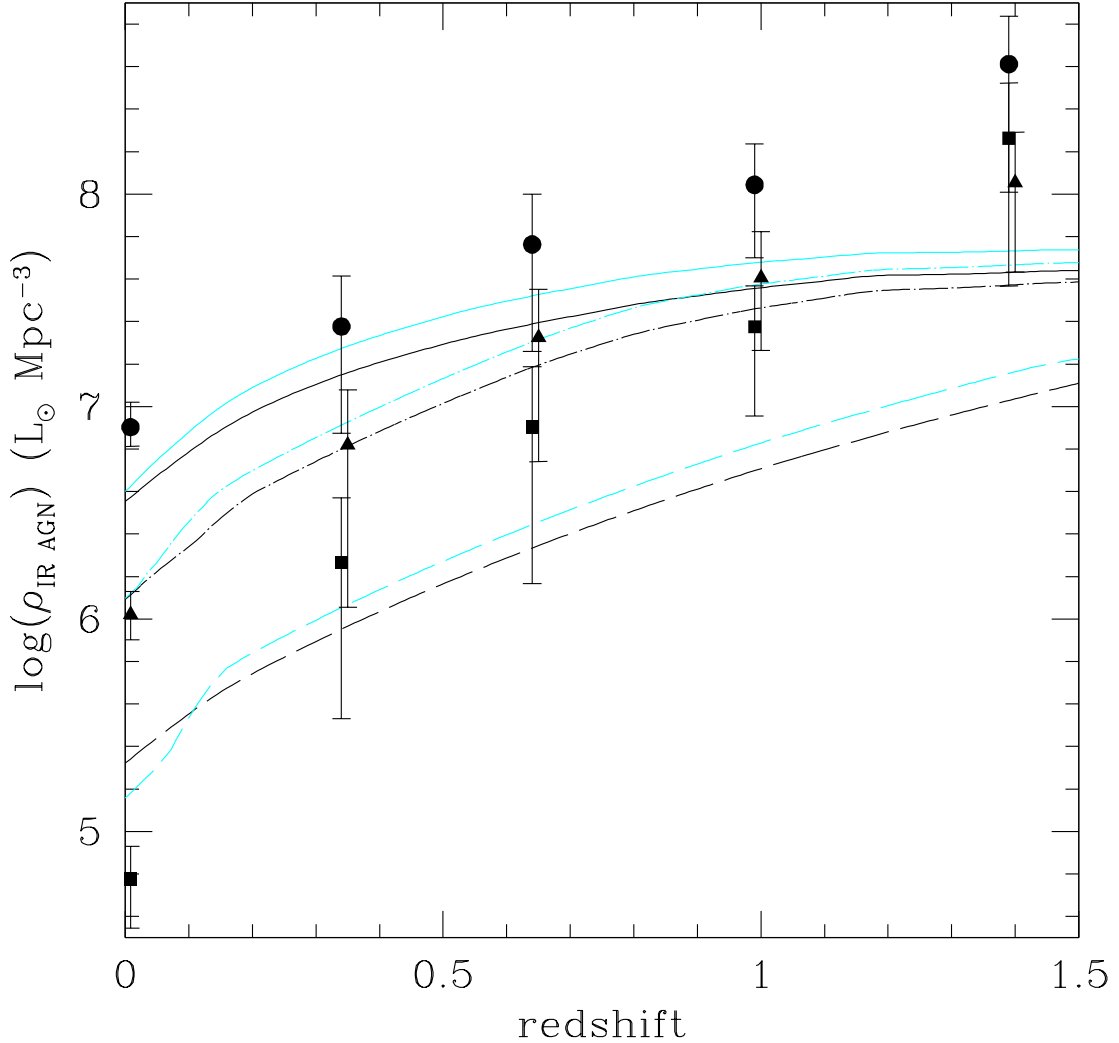}
\end{center}
\caption{Infrared luminosity density of bare AGN with respect to redshift.  Solid lines and circles refer to $\rho_{IR}$, dot-dashed lines and triangles refer to $\rho_{IR}^{LIRG}$, while dashed lines and squares refer to $\rho_{IR}^{ULIRG}$.  The black lines show the composite model and the cyan lines show the original model.  The $z$ = 0.0082 data points are taken from \citet{G10b} and the higher redshift data points are from the work of \citet{G10a} in the AKARI NEP deep field.}
\label{fig:lumden}
\end{figure*}

The cosmic infrared luminosity density and its evolution are good indicators of the star formation rate density 
and the evolution of star formation in the universe.  However, infrared luminosity density is significantly 
contaminated by AGN, thus making it important to understand the contribution of AGN to the infrared luminosity 
density and how the AGN contribution to the infrared luminosity density evolves.  The 8--1000 $\mu$m AGN luminosity density, 
$\rho_{IR}$, is calculated as
\begin{equation}
\rho_{IR}(z)= \int_{L_{IR, min}}^{L_{IR, max}} L_{IR} \frac{d\Phi_{\lambda} (L_{IR}, z)}{d\log L_{IR}} d\log L_{IR},
\label{eq:lumden}
\end{equation}
where $d\Phi_{\lambda}/d\log L_{IR}$ is the evolving Eddington ratio space density calculated in \citet{DB10} 
in terms of the total infrared luminosity, i.e. a total infrared luminosity function for a population of AGN with a 
specific Eddington ratio, which is found using the relation
\begin{equation}
\frac{d\Phi_{\lambda} (L_{IR}, z)}{d(\log L_{IR})} = \frac{d\Phi_{\lambda} (L_{X}, z)}{d(\log L_{X})} \frac{d(\log L_X)}{d(\log L_{IR})}.
\label{eq:irlf}
\end{equation}
As seen in figure \ref{fig:lumden}, both the composite and original models are in agreement with 
observations by \citet{G10a} at $z$ $\approx$ 0.2--0.8.  For $z$ $\gtrsim$ 1 both models do not 
evolve fast enough to agree with the $\rho_{IR}$ and ULIRG AGN infrared luminosity density, $\rho_{IR}^{ULIRG}$. However, both 
models are in fairly good agreement with the high redshift LIRG AGN luminosity density, $\rho_{IR}^{LIRG}$.  Locally 
both the original and composite model under-predict the $\rho_{IR}$ and 
over-predict the $\rho_{IR}^{ULIRG}$, but are in agreement with the $\rho_{IR}^{LIRG}$, 
as measured by \citet{G10b}.  Due to the method used to separate the AGN 
and star forming galaxies in the samples of \citet{G10a,G10b}, there is an uncertainty of up 
to 50$\%$ due to the complication of galaxies whose infrared SED is not clearly dominated by AGN 
activity nor by star formation \citep{G10b}.  Also, by applying the same method of separating the 
contribution of AGN and star forming galaxies to the infrared luminosity density measured by \citet{Sey10} using {\em Spitzer} 
24 $\mu$m sources in the CDF-S, the local $\rho_{IR}^{ULIRG}$ is $\sim$0.4 dex higher 
and in good agreement with local $\rho_{IR}^{ULIRG}$ predicted here.


\section{Accounting for Star Formation}
\label{sect:sf}

The infrared SEDs of most AGN are dominated by star formation \citep[e.g.,][]{F10}.  Unless the AGN host 
galaxy can be spatially resolved, it is very difficult to separate the AGN emission from the star 
formation emission.  Therefore it is important to consider how various AGN star 
formation scenarios will affect the predictions discussed above.  In particular, a constant star 
formation rate is considered, keeping with the unified scheme.  A star formation scenario consistent 
with the AGN evolution scenario discussed by \citet{S88} is also analyzed.  Finally, star formation 
scenarios which include evolution of the SFR with redshift and AGN $L_X$ are considered.

To calculate the star formation SED for a given SFR, a modified black body spectrum characterized 
by the dust emissivity, $\beta$, and the dust temperate, $T_d$, is used.  
The \citet{K98} relation,
\begin{equation}
SFR = 4.5\times10^{-44}L_{IR},
\label{eq:lir}
\end{equation}
is used to determine $L_{IR}$, in erg s$^{-1}$, for a given SFR, in M$_{\odot}$ yr$^{-1}$, and the $L_{IR}$-$T_d$ relation from \citet{Am10},
\begin{equation}
T_d = T_0 + \alpha\log(L_{IR}/L_{\odot}),
\label{eq:Td}
\end{equation}
is used to determine $T_d$.  Using {\em Herschel} observations, \citet{Am10} find 
$T_0$ = 20.5 K and $\alpha$ = 4.4, using $\beta$ = 1.5.  \citet{Am10} define $L_{IR}$ as the 8--1100 
$\mu$m luminosity, instead of the 8--1000 $\mu$m luminosity as done here; however, for our purposes, 
the difference is negligible.  The resulting star formation SEDs are in good agreement with the
star formation templates presented by \citet{Ri09} at rest frame 
wavelengths greater than 50 $\mu$m. The star formation SEDs are then added to the model SEDs 
discussed above to create the AGN+SF SED.  The calculations discussed in section \S \ref{sect:bare} 
are repeated using the AGN+SF SEDs in place of the AGN model SEDs.  Here, the predictions for the 
composite model are presented.  A comparison of the predictions of the original and composite models is discussed in \S \ref{sub:diff}.

\subsection{Constant Star Formation}
\label{sub:constant}
\begin{figure*}
\begin{center}
\includegraphics[angle=0, width=0.95\textwidth]{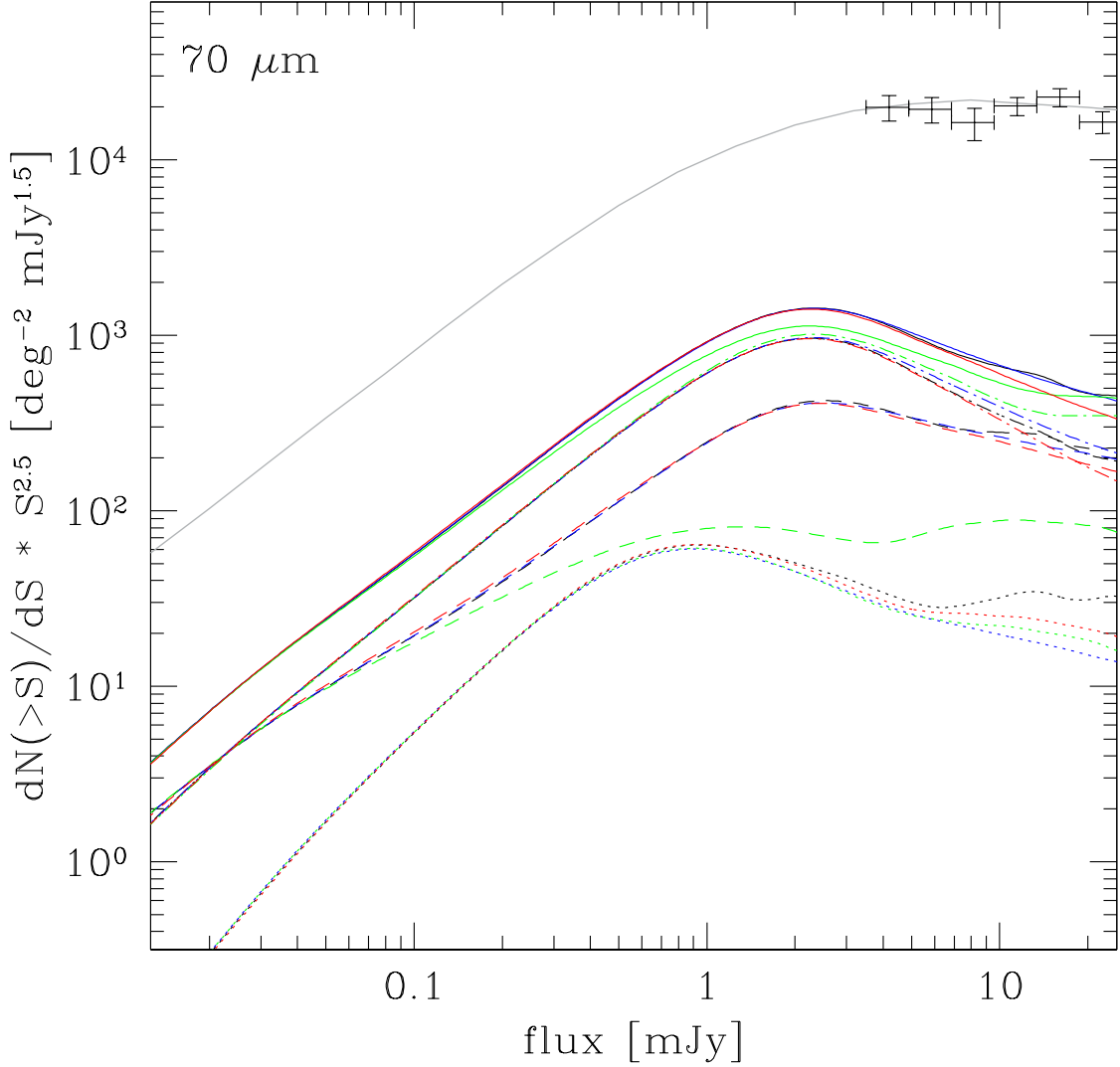}
\end{center}
\caption{Euclidean normalized differential number counts for AGN and host star formation at 70 $\mu$m for the various star formation scenarios using the composite model.  Constant star formation is shown as black.  The AGN evolution star formation scenario is shown as green.  Star formation with the redshift evolution found by \citet{S10} is shown as red.  The star formation scenario using the redshift and AGN $L_X$ evolution by \citet{W10} is shown in blue.   The line styles are the same as in figure \ref{fig:baredNdS}.  The grey line shows the best fit model galaxy differential number counts of \citet{F10}. {\em Spitzer} data points are from \citet{Beth10}.}
\label{fig:dnds70}
\end{figure*}

\begin{figure*}
\begin{center}
\includegraphics[angle=0, width=0.95\textwidth]{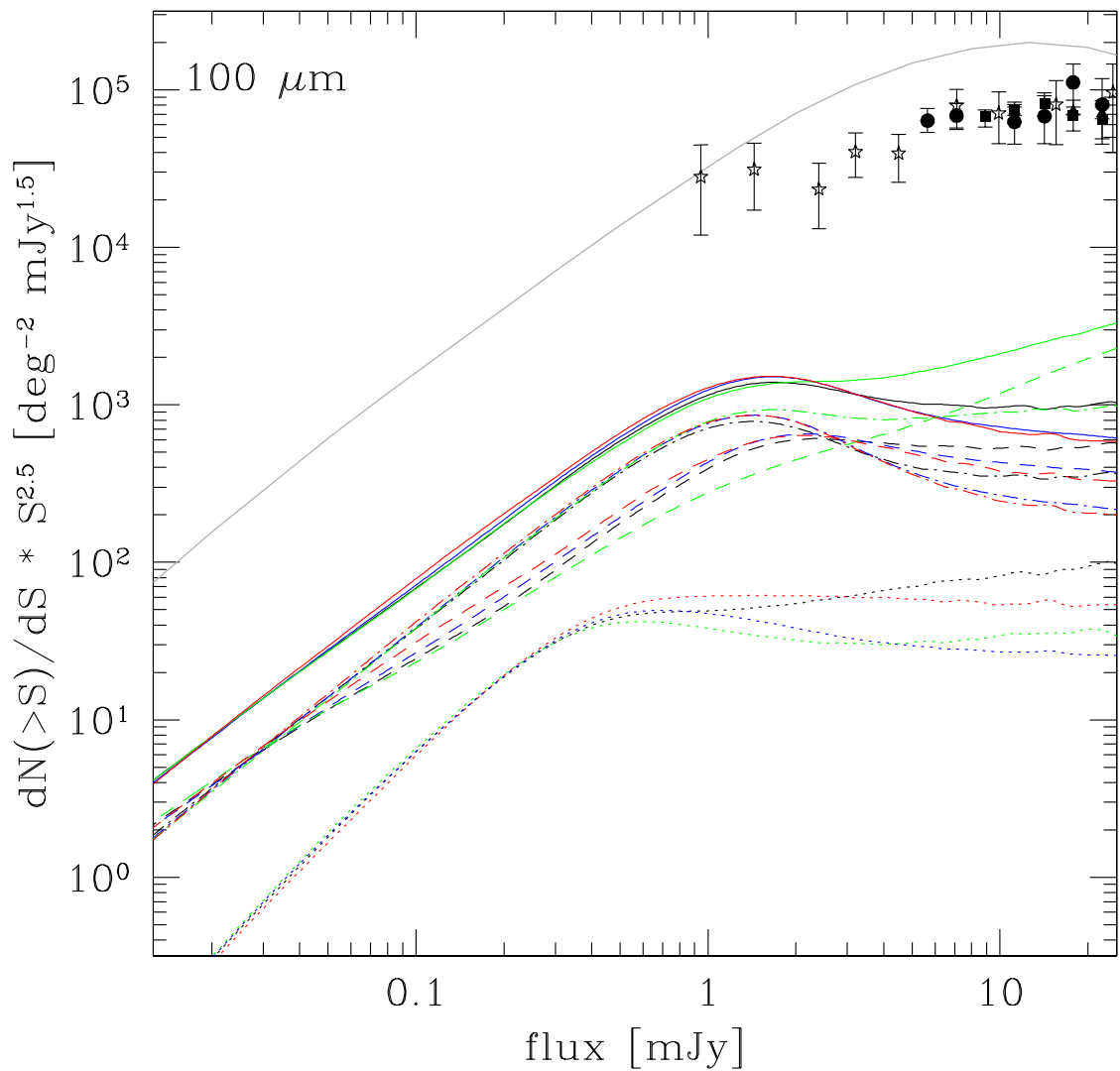}
\end{center}
\caption{Euclidean normalized differential number counts for AGN and host star formation at 100 $\mu$m for the various star formation scenarios using the composite model.  Lines are the same as in figure \ref{fig:dnds70}.  {\em Herschel} data points are from \citet{A10} (stars) and \citet{B10} (circles--GOODS-N, squares--Lockman XMM, and triangles--COSMOS).}
\label{fig:dnds100}
\end{figure*}

\begin{figure*}
\begin{center}
\includegraphics[angle=0, width=0.95\textwidth]{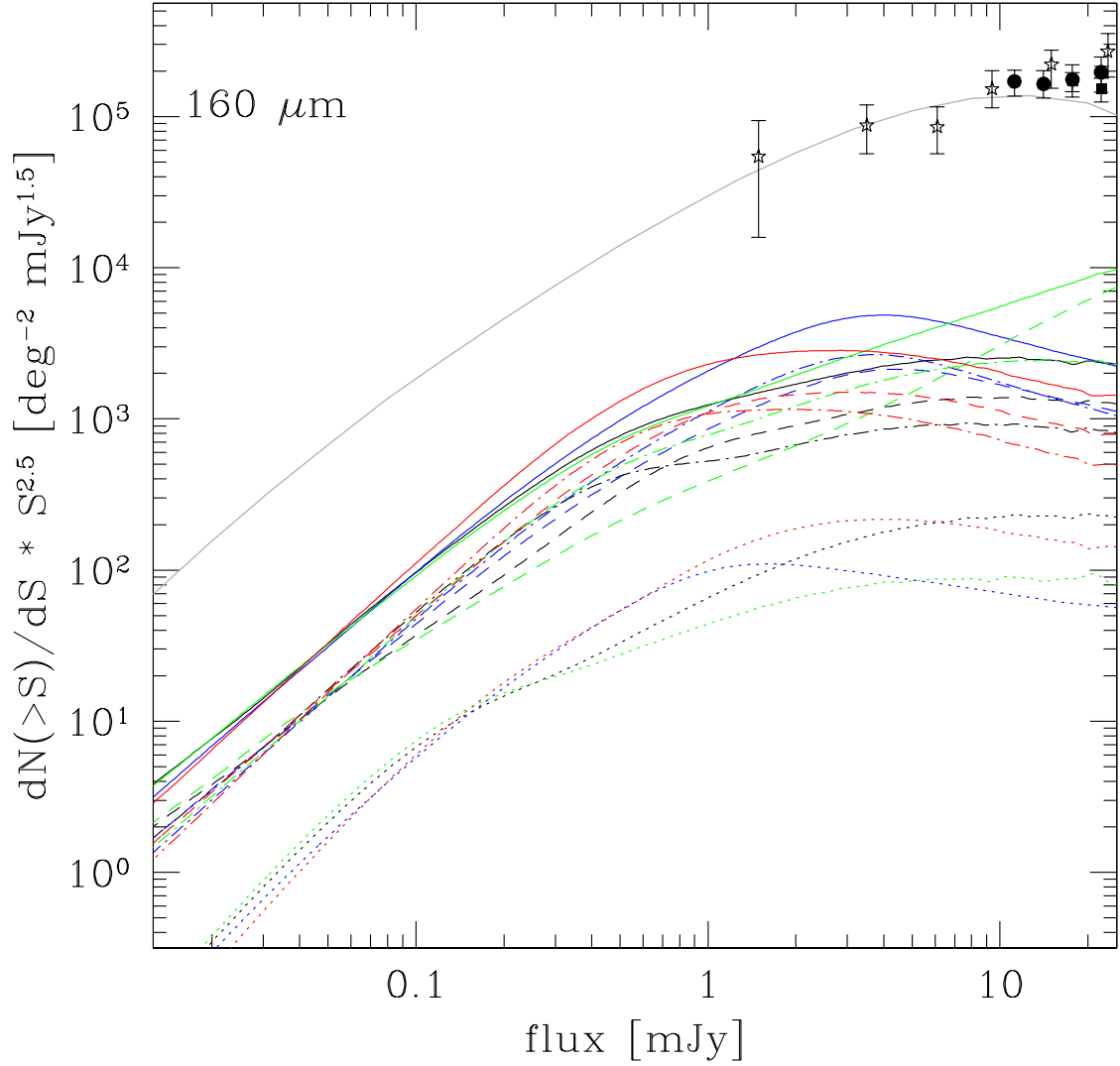}
\end{center}
\caption{Euclidean normalized differential number counts for AGN and host star formation at 160 $\mu$m for the various star formation scenarios using the composite model.  Line colors and styles are the same as in figure \ref{fig:dnds70}.  Data points are the same as in figure \ref{fig:dnds100}.}
\label{fig:dnds160}
\end{figure*}

\begin{figure*}
\begin{center}
\includegraphics[angle=0, width=0.95\textwidth]{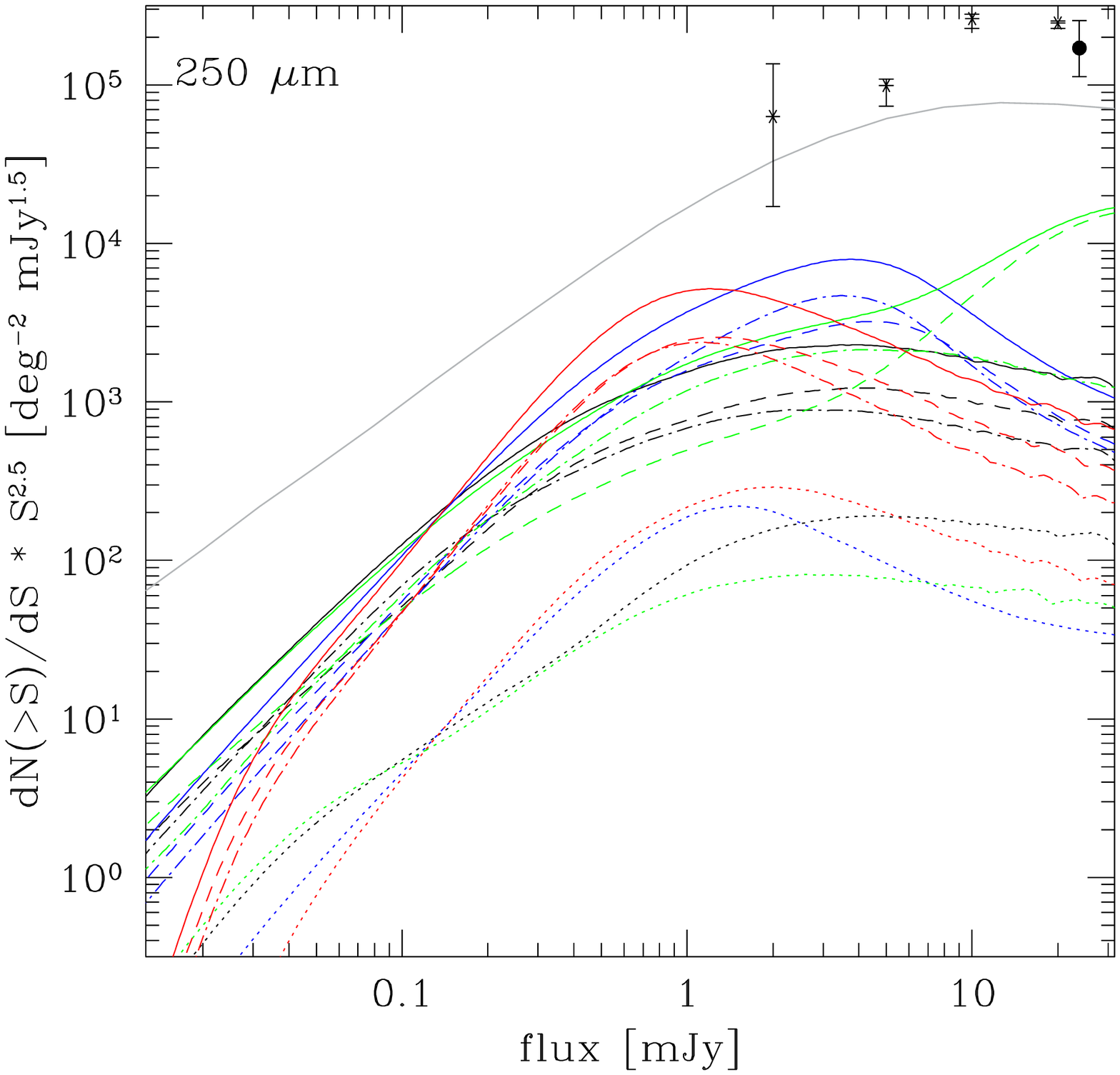}
\end{center}
\caption{Euclidean normalized differential number counts for AGN and host star formation at 250 $\mu$m for the various star formation scenarios using the composite model.  Line colors and styles are the same as in figure \ref{fig:dnds70}.  Circles are data points from {\em Herschel} \citep{O10} and astrisks show the multiply-broken power-law model of \citet{Glenn10}.}
\label{fig:dnds250}
\end{figure*}

\begin{figure*}
\begin{center}
\includegraphics[angle=0, width=0.95\textwidth]{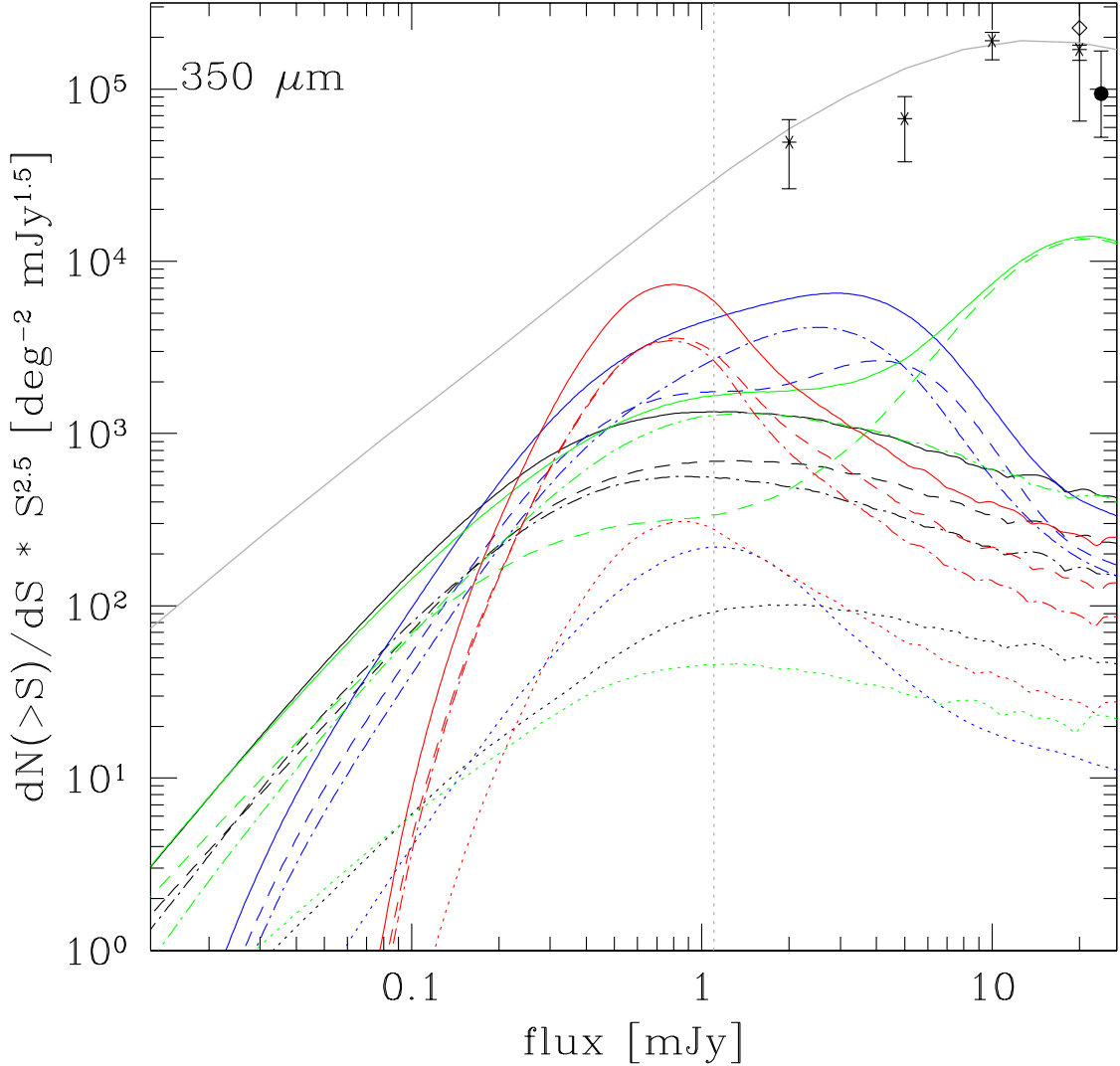}
\end{center}
\caption{Euclidean normalized differential number counts for AGN and host star formation at 350 $\mu$m for the various star formation scenarios using the composite model.  Line colors and styles are the same as in figure \ref{fig:dnds70}.  Additionally, the dotted grey line shows the expected continuum sensitivity of ALMA for an integration time of 60 seconds and a spectral resolution of 1 km/s.  Circles are data points from {\em Herschel} \citep{O10}, the diamond is from SHARC II \citep{K07}, and astrisks show the multiply-broken power-law model of \citet{Glenn10}.}
\label{fig:dnds350}
\end{figure*}

\begin{figure*}
\begin{center}
\includegraphics[angle=0, width=0.95\textwidth]{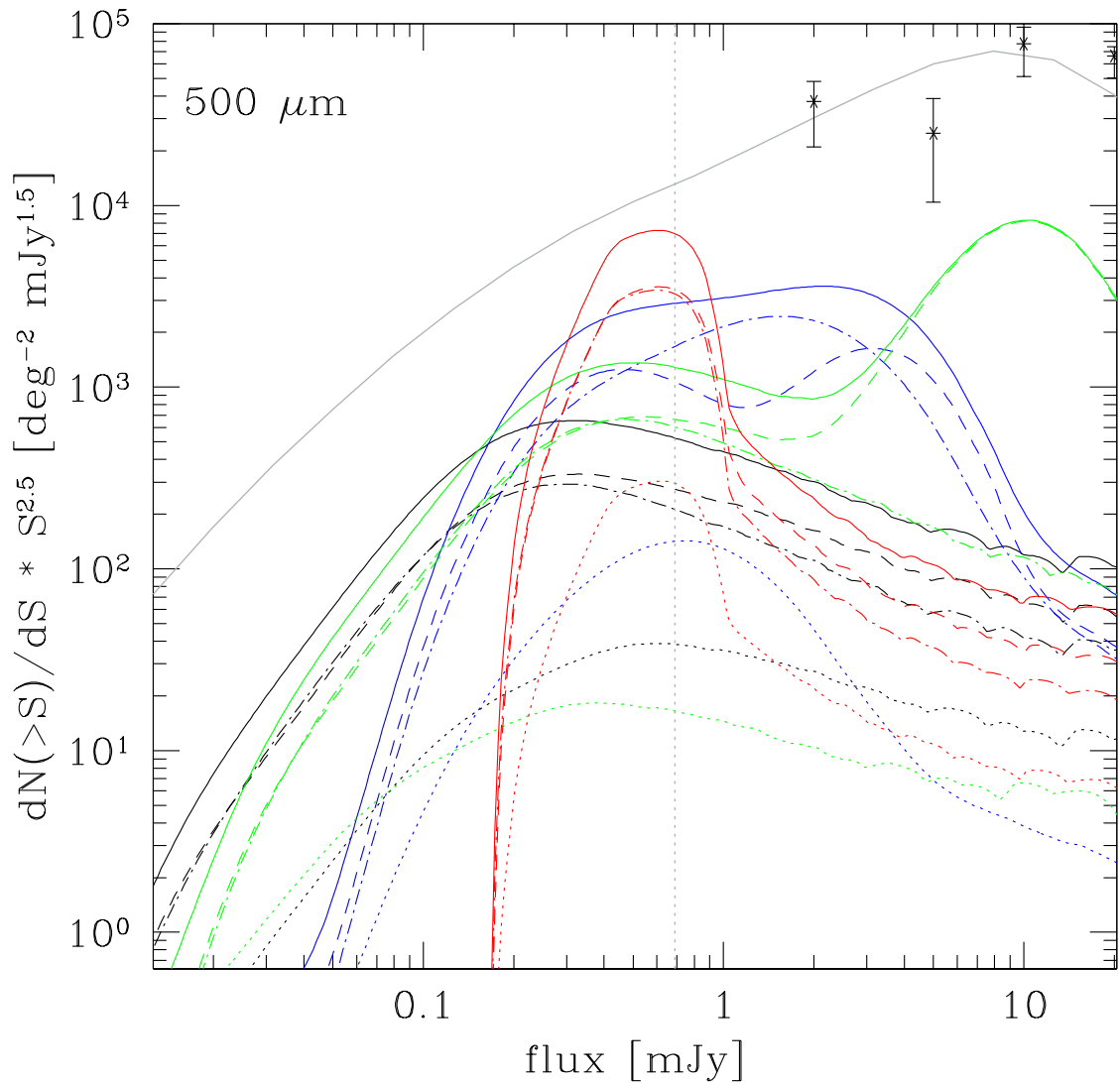}
\end{center}
\caption{Euclidean normalized differential number counts for AGN and host star formation at 500 $\mu$m for the various star formation scenarios using the composite model.  Line colors and styles are the same as in figure \ref{fig:dnds350}.  Circles are data points from {\em Herschel} \citep{O10} and astrisks show the multiply-broken power-law model of \citet{Glenn10}.  At bright fluxes numerical artifacts are present due to the small number of sources in this flux region.}
\label{fig:dnds500}
\end{figure*}

The simplest star formation scenario is where AGN hosts have a constant SFR.  \citet{BP07} 
found that an average AGN SFR = 1.0 M$_{\odot}$ yr$^{-1}$ reproduces the AGN contribution 
to the mid-infrared portion of the CIRB as measured by {\em Spitzer}; thus for the constant star 
formation model we set SFR = 1.0 M$_{\odot}$ yr$^{-1}$.  In keeping with the unified 
scheme, all AGN have the same SFR regardless of 
spectroscopic type.  The Euclidean normalized differential number counts for the constant star 
formation model are shown as the black lines in figures \ref{fig:dnds70}, \ref{fig:dnds100}, \ref{fig:dnds160}, 
\ref{fig:dnds250}, \ref{fig:dnds350}, and \ref{fig:dnds500}.  At wavelengths shorter than 350 $\mu$m, the number 
counts increase steeply from 0.01 mJy to $\sim$1 mJy and then remain approximately flat through 
$\sim$25 mJy.  At wavelengths longer than 350 $\mu$m, the number counts turn over around 1 
mJy and continue to decrease toward brighter flux levels.  At all wavelengths the CT AGN 
dominate the bright end counts above $\sim$1 mJy, with type 2 AGN dominating at lower flux 
levels.

\begin{figure*}
\begin{center}
\includegraphics[angle=0, width=0.95\textwidth]{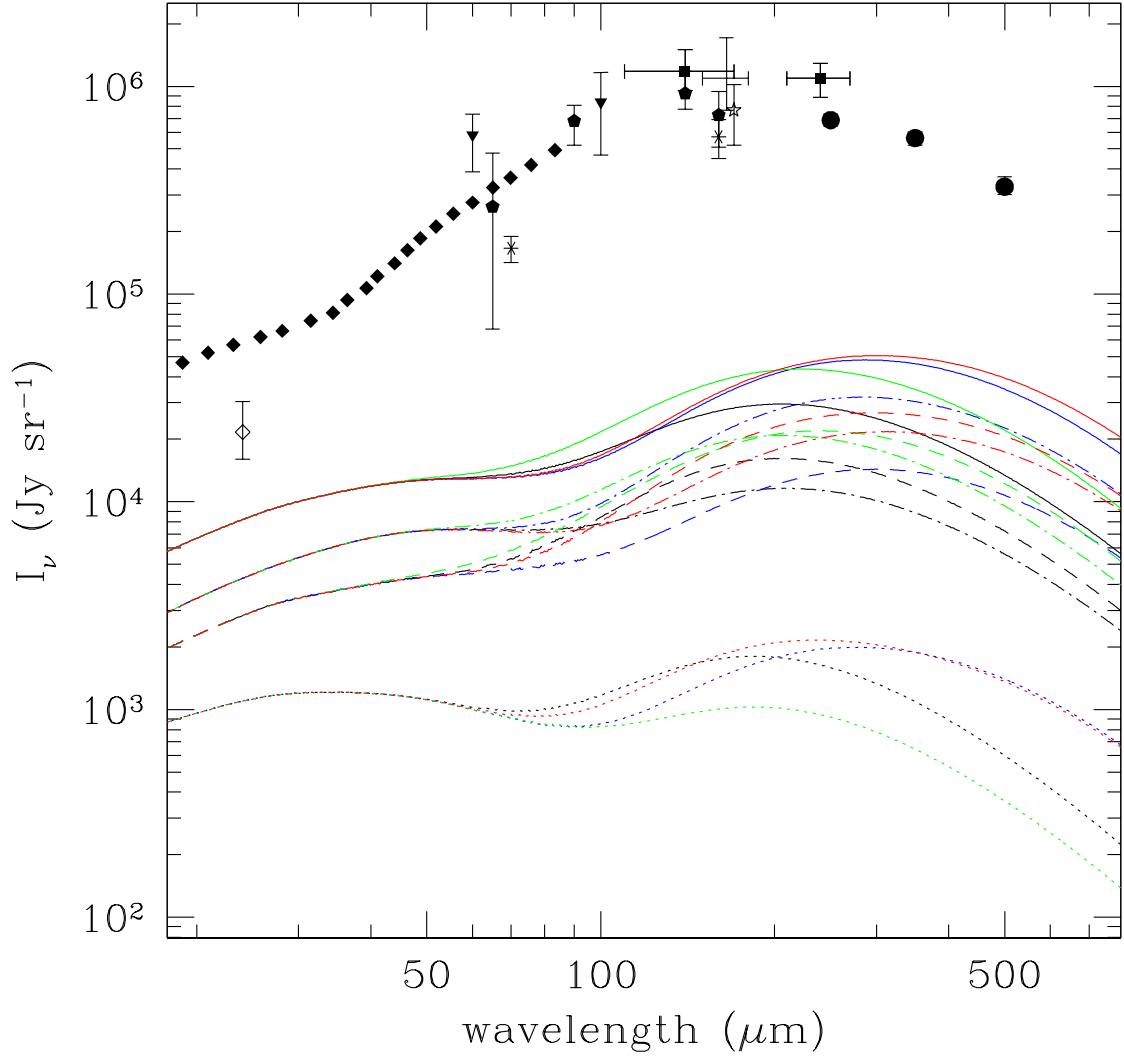}
\end{center}
\caption{AGN and host star formation contribution to the CIRB for the various star formation scenarios using the composite model.  Line colors are the same as in figure \ref{fig:dnds100}.  The line styles are the same as in figure \ref{fig:baredNdS}.  Data points are the same as figure \ref{fig:cirb}.}
\label{fig:cirbsf}
\end{figure*}

The black lines in figure \ref{fig:cirbsf} show the AGN contribution to the CIRB for the constant 
star formation model.  In this star formation scenario, AGN contribute $\sim$4$\%$ of the CIRB at 
160 $\mu$m.  The AGN contribution to the CIRB is dominated by CT AGN above 100 $\mu$m and type 2 
AGN also contribute significantly.  The AGN contribution to the CIRB peaks at approximately the 
same wavelength as the CIRB itself peaks, $\sim$160 $\mu$m.

\begin{figure*}
\begin{center}
\includegraphics[angle=0, width=0.95\textwidth]{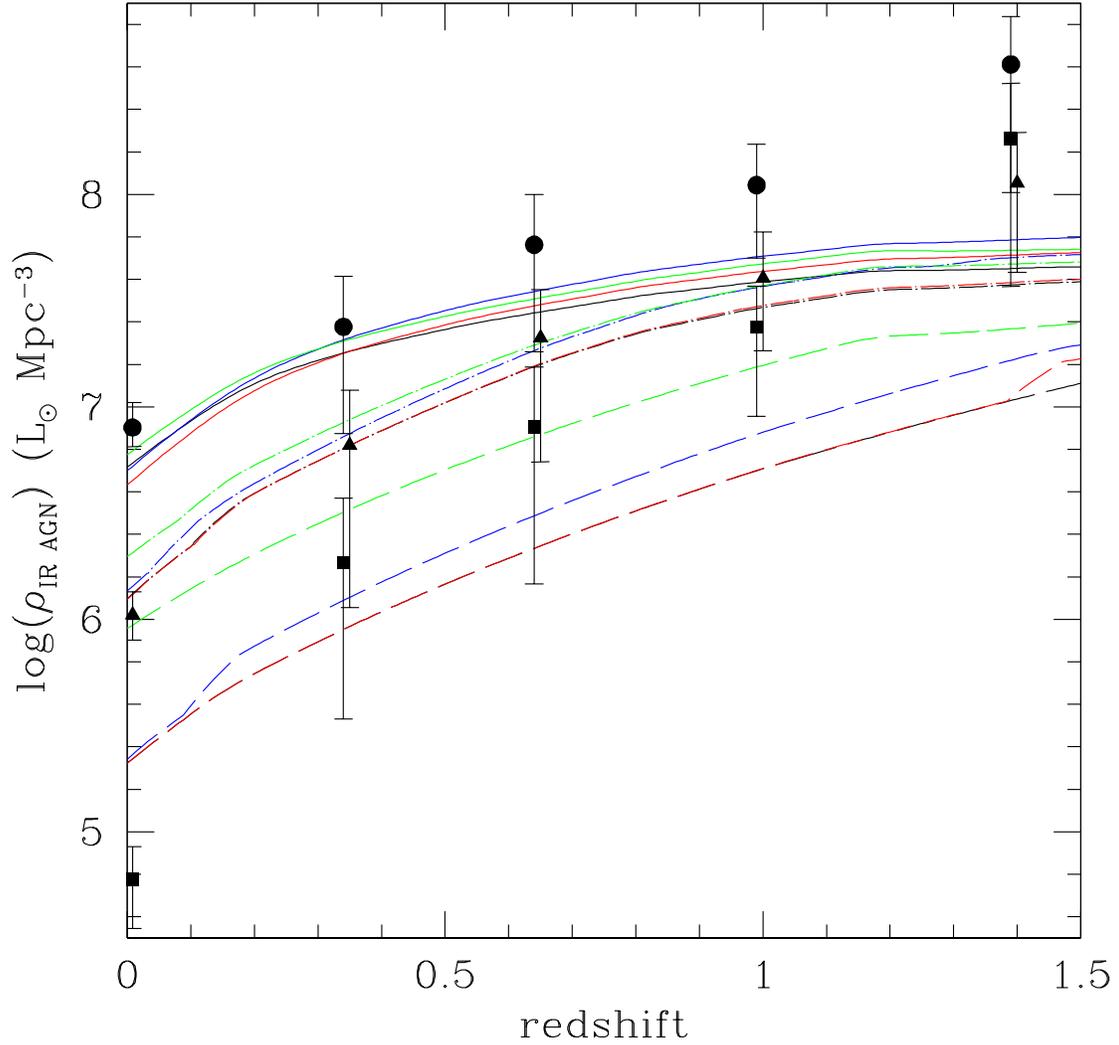}
\end{center}
\caption{AGN and host star formation infrared luminosity density for the various star formation scenarios using the composite model.  Line colors are the same as in figure \ref{fig:dnds100}.  The line styles and data points are the same as figure \ref{fig:lumden}. }
\label{fig:lumdensf}
\end{figure*}

The $\rho_{IR}$ is shown in figure \ref{fig:lumdensf}, with the constant star 
formation model shown as the black lines.  The $\rho_{IR}^{ULIRG}$ is not increased 
over the bare AGN scenario, however, the total and LIRG AGN infrared luminosity densities are increased 
slightly.  Thus the constant star formation model does not allow for rapid enough evolution with 
redshift to match the observed high redshift $\rho_{IR}$.

\subsection{AGN Evolution Scenario}
\label{sub:evol}
It has been suggested that, at least in the quasar regime, the level of obscuration observed in 
an AGN is connected to the evolutionary stage of the quasar \citep[e.g.,][]{S88, P04, B08, DB10}.  
Galaxy merger simulations support the evolutionary scenario showing that gas rich mergers lead to a burst of star formation 
and intense, highly obscured black hole growth \citep[e.g.,][]{H06}.  In the AGN evolution 
scenario, high Eddington ratio CT AGN would be expected to have SFRs reaching into the ULIRG 
regime and type 2 AGN hosts would be expected to have more star formation that type 1 AGN hosts 
\citep[e.g.,][]{P04}.  Following this prescription we set the high Eddington ratio SFR = 175 
M$_{\odot}$ yr$^{-1}$, the type 2 SFR = 2.0 M$_{\odot}$ yr$^{-1}$, and type 1 SFR = 0.5 M$_{\odot}$ 
yr$^{-1}$.  The low Eddington ratio CT AGN are also given SFR = 0.5 M$_{\odot}$ yr$^{-1}$ as these 
are weak AGN most likely obscured by molecular clouds in the host bulge or by dust lanes in the 
host galaxy \citep{MS09}, and not by intense nuclear starbursts.  This star formation scenario 
gives predictions which are only negligibly different from the scenario where all AGN have SFR = 
1.0 M$_{\odot}$ yr$^{-1}$ except for the high Eddington ratio CT AGN which have SFR = 175 
M$_{\odot}$ yr$^{-1}$.

The Euclidean normalized differential counts for the AGN evolution scenario are shown as 
the green lines in figures \ref{fig:dnds70}, \ref{fig:dnds100}, \ref{fig:dnds160}, \ref{fig:dnds250}, 
\ref{fig:dnds350}, and \ref{fig:dnds500}.  At all wavelengths the number counts rise as the flux level increases, 
leveling off around 1 mJy, and then continue to rise at least until the flux level of 10 mJy.  
CT AGN dominate at flux levels greater than $\sim$6 mJy and type 2 AGN dominate at lower flux 
levels for all wavelengths.

In figure \ref{fig:cirbsf}, the AGN contribution to the CIRB for the AGN evolution scenario is 
shown in green.  At 160 $\mu$m, AGN contribute $\sim$6$\%$ of the CIRB.  CT AGN dominate the AGN 
contribution to the CIRB at wavelengths greater than $\sim$200 $\mu$m and type 2 AGN dominate 
below 200 $\mu$m.  The peak of the AGN contribution to the CIRB is roughly at the same wavelength 
as the peak of the CIRB.

The green lines in figure \ref{fig:lumdensf} show the $\rho_{IR}$ for the AGN 
evolution scenario.  The total $\rho_{IR}$ is in decent agreement with observations 
except at the highest redshift bin, which this model under-predicts.  The local $\rho_{IR}^{LIRG}$ and $\rho_{IR}^{ULIRG}$ 
are over-predicted by this model but are in good agreement with observations at higher redshifts.  
However, the $\rho_{IR}^{ULIRG}$ observations show a stronger redshift evolution than predicted by 
this model.

\subsection{Evolution with Redshift}
\label{evolz}
Also considered was the star formation redshift evolution found by \citet{S10}, where SFR $\propto$ (1.0+$z$)$^{2.3}$.  
\citet{S10} did find that the highest luminosity quasars had a much stronger redshift evolution, perhaps as strong 
as SFR $\propto$ (1.0+$z$)$^{10}$.  For simplicity the total volume-averaged star formation rate redshift evolution 
is used here, independent of the object luminosity.  The local star formation rate is set such that SFR($z$=0.0) = 
0.5 M$_{\odot}$ yr$^{-1}$, in keeping with the average type 1 SFR found in a sample of local SDSS quasars by \citet{K06}. 
Following the unified scheme, the SFR is not dependent on spectroscopic type.  

As shown by the red lines in figures \ref{fig:dnds70}, \ref{fig:dnds100}, \ref{fig:dnds160}, \ref{fig:dnds250}, \ref{fig:dnds350}, and \ref{fig:dnds500}, 
the Euclidean normalized differential number counts increase with increasing flux level until peaking at $\sim$1 mJy.  
The number counts decrease at flux levels above $\sim$1 mJy, and, for wavelengths greater than 160 $\mu$m, the number 
counts decrease more steeply as the wavelength increases. CT AGN dominate the counts on the brighter side of the peak 
and type 2 AGN dominate the lower flux level side of the peak.

The star formation redshift evolution model contribution to the CIRB is shown in red in figure \ref{fig:cirbsf}.  The 
AGN contribution to the CIRB at 160$\mu$m is $\sim$5$\%$.  CT AGN dominate the AGN contribution to the CIRB at wavelengths 
greater than $\sim$100 $\mu$m, with type 2 AGN dominating at lower wavelengths.  The AGN contribution to the CIRB for 
this model peaks at $\sim$320 $\mu$m, a significantly longer wavelength than the peak of the CIRB as a whole.

The $\rho_{IR}^{LIRG}$ and $\rho_{IR}^{ULIRG}$ for the star formation redshift evolution model are very similar to 
those for the constant star formation scenario, as shown by the red lines in figure \ref{fig:lumdensf}.  The 
total $\rho_{IR}$  of the star formation redshift evolution model under-predicts the observed 
local $\rho_{IR}$ but evolves more strongly with redshift than the constant star formation scenario.

\subsection{Evolution with Redshift and AGN $L_X$}
\label{sub:evolzlx}
The final star formation scenario investigated here is the redshift and AGN $L_X$ dependent 
star formation scenario used by \citet{W10}, where 
\begin{equation}
SFR \propto\sqrt{L_X/10^{43}}(1.0+z)^{1.6}.
\label{eq:sfr1}
\end{equation}
Type 1 AGN are given the normalization constant 0.63 M$_{\odot}$ yr$^{-1}$ and for type 2 
AGN the normalization prefactor is increased to 2.0 M$_{\odot}$ yr$^{-1}$ \citep{W10}.  
High Eddington ratio CT AGN are given the type 2 SFR and low Eddington ratio CT AGN are 
given the type 1 SFR, for the reasons discussed in \S \ref{sub:evol}.

The Euclidean normalized differential number counts for the redshift and AGN $L_X$ 
dependent SFR model are shown as blue lines in figures \ref{fig:dnds70},
\ref{fig:dnds100}, \ref{fig:dnds160}, \ref{fig:dnds250}, \ref{fig:dnds350}, and \ref{fig:dnds500} and are in decent 
agreement with the predictions made by \citet{W10} based on a simulation of the extragalactic radio sky.  The 
differential number counts increase with increasing flux level until peaking at 1--3 mJy, depending 
on wavelength.  The peak of the number counts appears to increase with wavelength, with 
the peak at $\sim$1 mJy at 100 $\mu$m and $\sim$3 mJy at 350 $\mu$m.  The number counts 
decrease from the peak to the brighter flux levels.  At 100, 350, and 500 $\mu$m CT 
AGN dominate the number counts on the brighter side of the peak, but at 70, 160, and 250 
$\mu$m type 2 AGN dominate except for at the brightest flux levels, $\gtrsim$10 mJy.

The blue lines in figure \ref{fig:cirbsf} show the AGN contribution to the CIRB for the 
redshift and AGN $L_X$ dependent SFR model.  This model predicts $\sim$5$\%$ of the CIRB 
at 160 $\mu$m is due to AGN.  Type 2 AGN dominate the AGN contribution to the CIRB at all 
wavelengths and CT AGN make a significant contribution.  For this model the AGN contribution 
to the CIRB peaks at $\sim$300 $\mu$m, which is significantly different from the peak of the 
total CIRB around 160 $\mu$m.

In figure \ref{fig:lumdensf} the $\rho_{IR}$ for the redshift and $L_X$ 
dependent SFR model is shown in blue.  This model is in decent agreement with the 
$\rho_{IR}^{LIRG}$ at all redshifts but over-predicts the local $\rho_{IR}^{ULIRG}$  
and under-predicts the local total $\rho_{IR}$.  Both the total $\rho_{IR}$ and $\rho_{IR}^{ULIRG}$ 
predictions are in agreement with observations at moderate redshifts, but do not 
evolve strongly enough with redshift to be in agreement with observations at the highest redshift.

\section{Discussion}
\label{sect:disc}

We have presented predictions for observations of AGN and AGN hosts in the FIR {\em Herschel} 
bands based on the composite model by \citet{DB10}\footnote{Models are available in table form by contacting the authors.}.  These findings demonstrate that while AGN 
may not contribute a large fraction of the CIRB, AGN will be significant FIR sources and 
care must be taken in FIR surveys to identify AGNs as such.  Here we discuss the implications of 
these results in terms of AGN and AGN host demographics.  

\subsection{CT AGN}
\label{sub:CT}

A substantial population of CT AGN are necessary for AGN population synthesis models to match 
the peak of the CXRB at $\sim$30 keV \citep[e.g.,][]{B06a, T09a, DB10}.  Due to the extreme 
levels of obscuration in CT AGN, these elusive sources are generally only observed in the 
very hard X-ray, $>$ 10 keV, or the infrared, especially the FIR.  Using X-ray 
stacking methods, it has been shown that a large fraction of bright infrared excess sources 
($f_{24}/f_R$ $>$ 1000) host heavily obscured AGN \citep[e.g.,][]{D07, F09, T09}.  However, 
it is uncertain how many of these highly obscured AGN are actually CT \citep[see][]{Gk10,Georg10}.  
Also, there is much debate over the prevalence of AGN in sources with more moderate infrared 
luminosities \citep[e.g.,][]{Dey08, D08, F09, T09, T10}.  Because CT AGN are generally not observable 
in the 2--10 keV band, AGN hard X-ray luminosity functions do not include CT AGN and therefore they 
must added in by hand to population synthesis models. In the population synthesis model used 
in this study, CT AGN are assumed to be accreting at either at $L/L_{Edd}$ $>$ 0.9 or $L/L_{Edd}$ $<$ 
0.01, with $f_{CT}$ independent of $f_2$, as found by \citet{DB10}.  Here we discuss predictions 
specifically for CT AGN in the FIR.  

At all wavelengths and for all star formation scenarios, the differential number counts are 
dominated by CT AGN for fluxes $\gtrsim$ 1--10 mJy.  Type 2 AGN dominate at lower fluxes.  Moreover, CT AGN, 
should constitute a non-trivial fraction of {\em Herschel} 
sources.  Depending on the star formation scenario considered, CT AGN could make up $\sim$10$\%$ of 
{\em Herschel} sources, even at 500 $\mu$m.

Depending on the star formation scenario, CT AGN are found to contribute $<$5$\%$ of the CIRB at 160 
$\mu$m.   For bare AGN, CT AGN dominate the AGN contribution to the CIRB at wavelengths $\gtrsim$ 
200 $\mu$m.  When star formation is included, CT AGN dominate the AGN contribution to the CIRB at 
wavelengths $\gtrsim$ 100 $\mu$m for all star formation scenarios, except the \citet{W10} model.  

Comparing the predictions made here against the infrared luminosity density found by \citet{Sey10}, 
we find that CT AGN and their hosts contribute $\sim$3$\%$ of the local infrared luminosity density 
from sources with $L_{IR}$ $>$ 10$^{10}$ L$_{\odot}$ and nearly one-fourth
of the infrared luminosity density from sources in the ULIRG range.  In the AGN evolution star 
formation scenario, CT AGN can account for all of the local ULIRG range luminosity density.  At $z\sim$ 1 the relative 
CT AGN contribution decreases significantly in all luminosity ranges.  However, when taking into account the stronger evolution 
of the SFR in high luminosity sources found by \citet{S10}, CT AGN and their hosts can still contribute 
nearly a quarter of the infrared luminosity density in the ULIRG range at $z$ $\sim$ 1.  Showing that at higher redshifts 
CT AGN contribute less to the total infrared luminosity density, but may still contribute quite significantly 
to the brightest sources.

The FIR is an important wavelength range for observing CT AGN due to the large amount of cold dust which 
obscures CT AGN.  The majority of AGN observed by {\em Herschel} will be CT.  Depending on the star 
formation trends in CT AGN hosts, CT AGN and their hosts may constitute nearly $\sim$10$\%$ of {\em Herschel} 
sources at 500 $\mu$m.  The relative contribution of CT AGN and their hosts to the ULIRG range infrared 
luminosity density is $\lesssim$25$\%$ and appears to be approximately constant over the redshift range $z$ = 0--1.  
However, \citet{Hat10} showed that AGN cannot be identified by their {\em Herschel}-SPIRE 
colors alone.  Therefore finding CT AGN in the FIR will require either {\em Spitzer}-MIPS coverage of bright SPIRE sources \citep{Hat10} 
or X-ray stacking.  Since the AGN and host differential number counts for both the composite and original models are 
dominated by CT AGN in the SPIRE bands, X-ray stacking of bright SPIRE sources is likely to disclose a large fraction of the CT AGN population.

\subsection{Differences Between Original and Composite Model}
\label{sub:diff}
The difference between the original model and the composite model is that in the composite model the 
CT AGN are put in specific, physically motivated Eddington ratio bins.  Also, in the original model $f_{CT}$ $\propto$ $f_2$, 
but in the composite model $f_{CT}$ is independent of $f_2$.   
In order to understand the effects of the differences between the two models, 
we compare the predictions of the original and composite model for the constant star formation scenario.  

In the differential number counts, the differences between the composite and original models are small 
but not insignificant.  At all wavelengths the original model has a steeper decline in the bright end 
counts than the composite model.  The original model predicts that the number counts will 
be dominated by type 2 AGN except for at the brightest fluxes.    Conversely, in the composite model the differential counts are dominated 
by CT AGN at every wavelength for fluxes $\gtrsim$ 1 mJy.  

The original model predicts a smaller overall AGN contribution to the CIRB than the composite model.  
In the composite model, CT AGN dominate the AGN 
contribution to the CIRB at wavelengths greater than 100 $\mu$m, but in the original model the type 2 AGN 
dominate the AGN contribution to the CIRB at all wavelengths.

The original and composite models make similar predictions as to the AGN contribution to the infrared 
luminosity density in all luminosity ranges.  The difference between these two models is most noticeable in the 
contribution of CT AGN to the local ULIRG range luminosity density.  The original model predicts that CT AGN 
contribute 4$\%$ of the local ULIRG infrared luminosity density.  The 
composite model predicts that nearly one-fourth of the local ULIRG infrared luminosity density is due to CT AGN.

The overall predictions of the original and composite models are in agreement.  To observationally determine which 
model best describes the CT AGN population, rigorous measurements of the CT AGN contribution to the local infrared 
luminosity density and/or accurate accounting of {\em Herschel} number counts to see whether CT AGN or type 2 AGN 
dominate the number counts will be needed.

\subsection{Star Formation in AGN hosts}
\label{sub:sf}

As seen in figures \ref{fig:dnds70}, \ref{fig:dnds100}, \ref{fig:dnds160}, \ref{fig:dnds250}, \ref{fig:dnds350}, and \ref{fig:dnds500},
the star formation scenario used greatly affects the predicted differential number counts.  At 
wavelengths $\lesssim$ 250 $\mu$m, the faint end slope is similar for all the star formation 
models investigated here, but the bright end slope is highly dependent on the host star formation 
at all wavelengths.  The differences between the various star formation scenarios becomes more 
prominent when observing at longer wavelengths.  The constant star formation model peaks at $\sim$ 1 
mJy and has a relatively flat bright end slope.  The AGN evolution  
star formation model peaks at $\gtrsim$ 10 mJy with a knee at $\sim$ 1 mJy.  
The redshift only evolution model peaks around, or just short of 1 mJy. This model also has the 
steepest faint and bright end slopes of the star formation scenarios considered here. 
The \citet{W10} star formation model peaks between 1 and 10 mJy for wavelengths $\gtrsim$ 100 $\mu$m.  The flux 
level of the peak in the differential counts may be an important tool in understanding the evolution of SFR in 
AGN hosts.  This tool will be most effective when observing at longer wavelengths.  


Dust obscured star formation is believed to be the primary progenitor of the CIRB with much debate as 
to the contribution from AGN.  \citet{BP07} found that AGN and star formation in AGN hosts can account for 
$\sim$30$\%$ of the CIRB at 70 $\mu$m.  Similarly, \citet{Mul10} find that AGN 
contribute 5--25$\%$ of the CIRB at 70 $\mu$m.  However, at longer wavelengths it appears the AGN contribution 
reduces to $\lesssim$10$\%$ \citep{J10,Lac10}. In this work it was found that AGN and host star formation 
contribute $\sim$5$\%$ of the CIRB at 160 $\mu$m.  However, when investigating the submillimeter properties of 
X-ray-selected AGN, \citet{L10} found the average AGN SFR to be $\sim$30 M$_{\odot}$ yr$^{-1}$, in 
which case AGN and their hosts would contribute $\sim$88$\%$ of the CIRB at 160 $\mu$m. Therefore, 
understanding the star formation trends in AGN hosts is necessary for understanding how significant 
the contribution of AGN and their hosts is.

The peak intensity of the CIRB occurs at $\sim$160 $\mu$m.  The star formation scenarios investigated 
here which take into account the evolution of the SFR with redshift predict the peak of the AGN contribution 
to the CIRB occurs at $\sim$300$\mu$m.  This suggests that the SFR of AGN hosts evolves differently with 
redshift than the SFR of normal galaxies.  This effect could also be explained if the star 
formation SEDs used here are on average too cold, which is unlikely.  If it is true that the SFR of AGN hosts evolves differently 
than the SFR of normal galaxies, this could offer important insights into the role of the AGN in the host galaxy evolution.
At wavelengths shorter than the peak of the CIRB, the dominate contribution is from sources $z$ $<$ 1.  Sources at 
$z$ $>$ 2 tend to dominate at wavelengths longer than 500 $\mu$m.

Another tool used to study different AGN host star formation scenarios is the infrared luminosity density. Using 
simulations, \citet{Hop10} find that AGN contribute 1-5$\%$ of the total infrared luminosity density at 
all redshifts.  Applying the classification scheme of \citet{Y10} to AKARI sources, \citet{G10b} find that AGN 
contribute $\sim$20$\%$ of the total infrared luminosity density, $\sim$40$\%$ of the luminosity density from sources 
in the LIRG range, and $\gtrsim$90$\%$ of the luminosity density of sources in the ULIRG range, regardless of redshift.  When considering the 
infrared luminosity density as measured by \citet{Sey10}, we find that the
AGN and host galaxy contribution to the local infrared luminosity density is approximately a factor of 2 smaller than that found by \citet{G10b}.   
At $z\sim$ 1 the relative AGN contribution decreases to $\lesssim$5$\%$ for all luminosity ranges.  
Even when taking into account the stronger evolution of the SFR in high luminosity sources found by \citet{S10}, 
it appears that AGN and their hosts only contribute 
$\sim$25$\%$ of the infrared luminosity density in the ULIRG luminosity range at $z$ $\sim$ 1.  The 
reduction in the AGN contribution to the ULIRG range infrared luminosity density by a factor of 2 between $z$ 
$\sim$ 0 and $z$ $\sim$ 1, is consistent with the findings of \citet{Sturm10} that ULIRG level luminosities can 
be achieved without major mergers at higher redshifts, suggesting that the AGN fraction in the high redshift 
ULIRG population will be smaller than that found locally.

Determining the star formation history of AGN hosts is important in understanding why some galaxies host active 
supermassive black holes (SMBHs) and other galaxies host inactive SMBHs.  The predictions presented here show that the flux 
level of the differential number counts peak in longer wavelength bands will be a helpful tool in determining 
the star formation history of AGN hosts.  By comparing the peak wavelength of the AGN contribution to the CIRB 
to the peak wavelength of the CIRB, it is possible to determine if the star formation evolution of AGN hosts is different 
from that of normal galaxies.  Also, at higher redshift AGN will have a smaller contribution to the ULIRG population, 
and that contribution will be dominated by CT AGN.

\subsection{Eddington Ratio Breakdown}
\label{sub:edd}

\begin{figure*}
\begin{center}
\includegraphics[angle=0, width=0.95\textwidth]{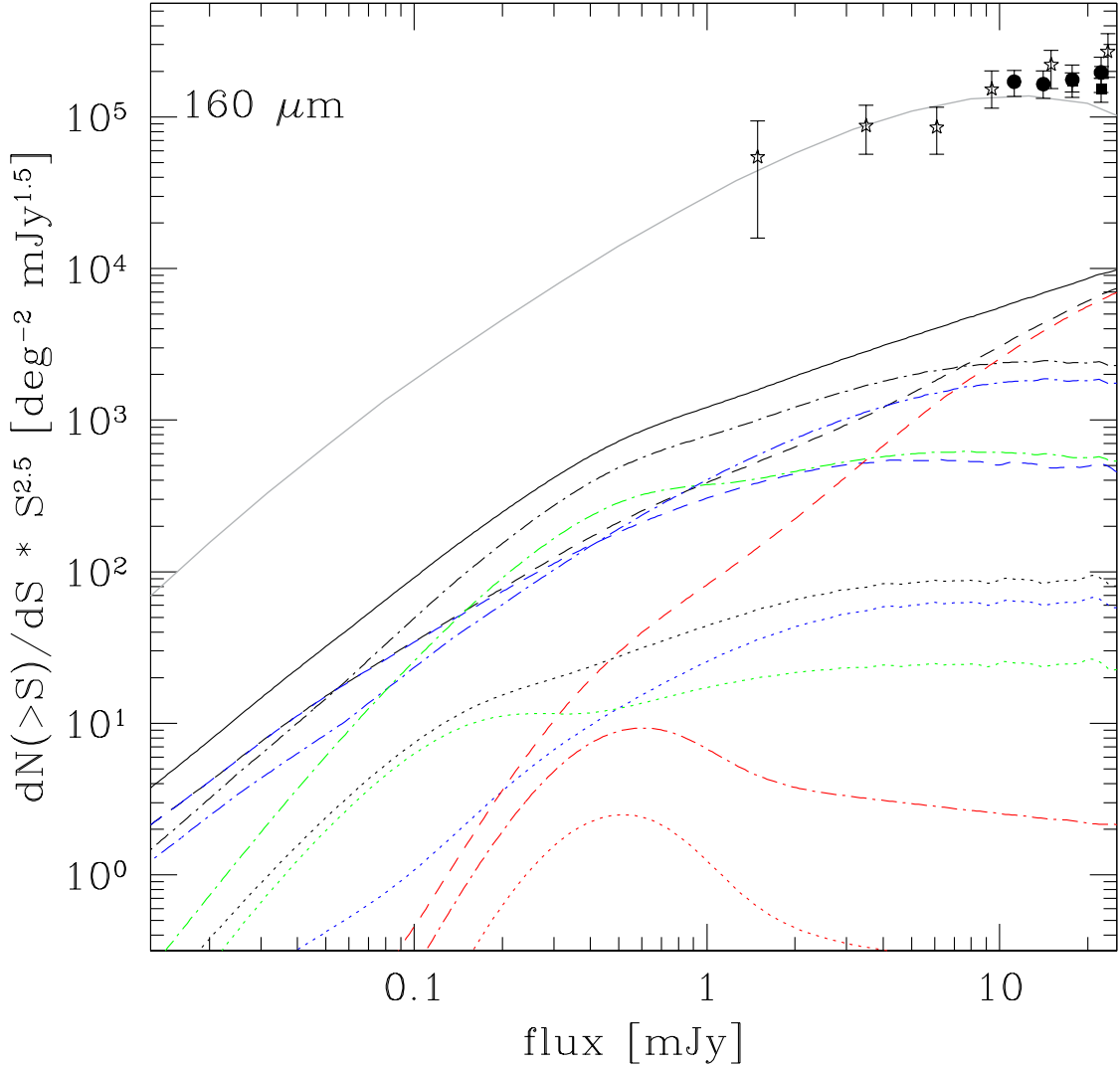}
\end{center}
\caption{Euclidean normalized differential number counts for AGN and host star formation at 160 $\mu$m for the AGN evolution star formation model.  Black lines show the total predictions for the composite model with the low Eddington ratio sources in blue, the mid Eddington ratio sources in green, and the high Eddington ratio sources in red.  The line styles are the same as in figure \ref{fig:baredNdS}.  The solid grey lines shows the best fit model galaxy differential number counts of \citet{F10}. the Data points are the same as in figure \ref{fig:dnds70}.}
\label{fig:edd160}
\end{figure*}

\begin{figure*}
\begin{center}
\includegraphics[angle=0, width=0.95\textwidth]{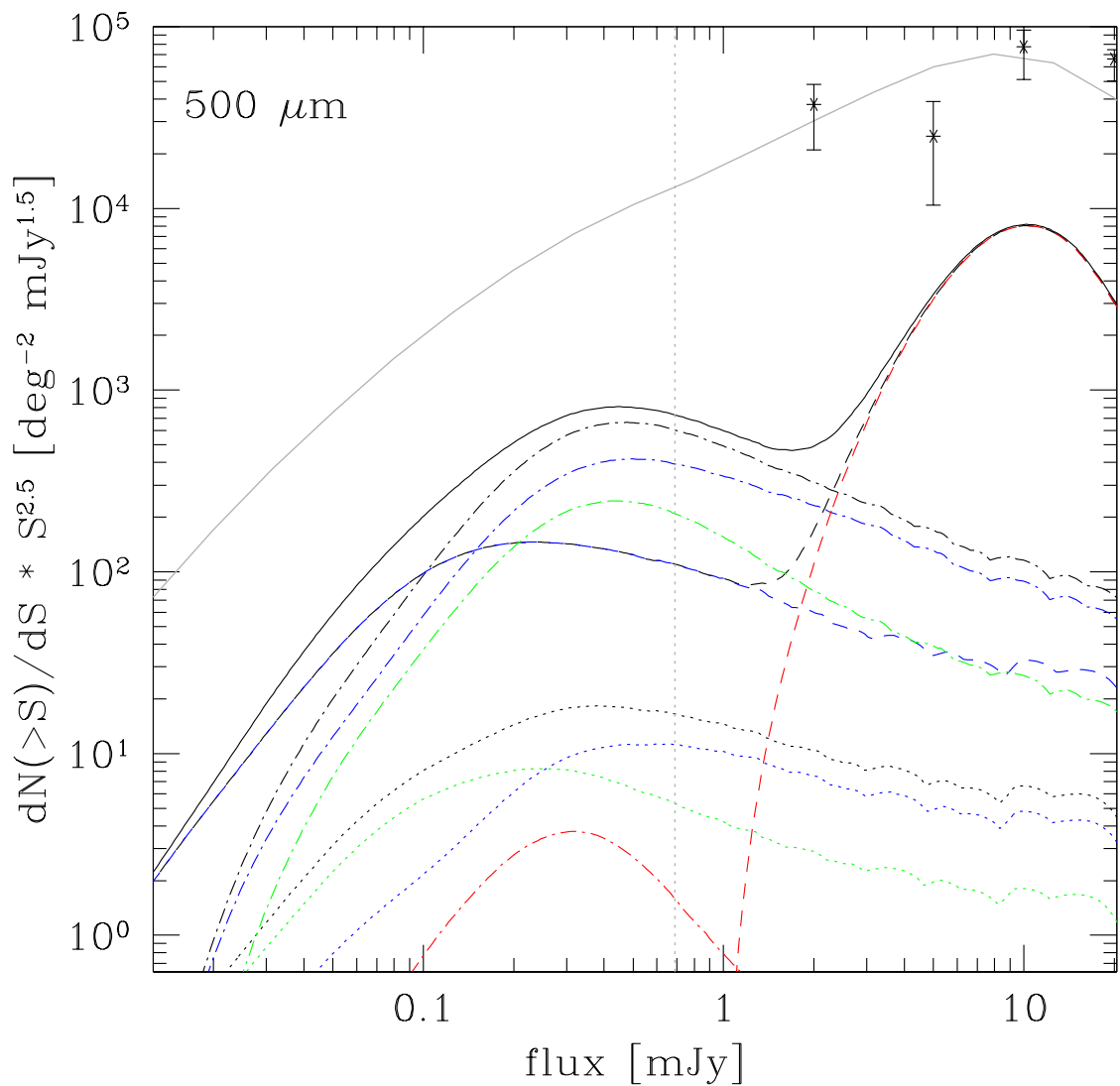}
\end{center}
\caption{Euclidean normalized differential number counts for AGN and host star formation at 160 $\mu$m for the AGN evolution star formation model.  Line colors and styles are the same as in figure \ref{fig:edd160} with the addition of the dotted grey line which shows the expected continuum sensitivity of ALMA for an integration time of 60 seconds and a spectral resolution of 1 km/s. Data points are the same as in figure \ref{fig:dnds250}.  At bright fluxes numerical artifacts are present due to the small number of sources in this flux region.}
\label{fig:edd500}
\end{figure*}

By using $d\Phi_{\lambda}(L_X,z)/d\log L_X$, the evolving Eddington ratio space density computed by \citet{DB10}, 
instead of a traditional luminosity function, it is possible to make predictions for the contribution of AGN with 
different Eddington ratios.  In figures \ref{fig:edd160} and \ref{fig:edd500} the composite model Euclidean normalized differential 
counts are shown for the AGN evolution star formation model at 160 and 500 $\mu$m with the relative contributions 
from the different Eddington ratio bins shown.  The blue lines show the contribution from AGN with $L/L_{Edd}$ $<$ 
0.01.  The green lines show the contribution from AGN with 0.01 $<$ $L/L_{Edd}$ $<$ 0.9.  The contribution from 
high Eddington ratio sources is shown in red.  For all but the AGN evolution scenario, the number counts are 
dominated by low Eddington ratio AGN at all flux levels, due to the high 
space density of low accretion rate AGN at all redshifts. As shown in figures \ref{fig:edd160} and \ref{fig:edd500}, 
the AGN evolutionary scenario bright end counts 
are dominated by high Eddington ratio CT AGN because of the high star formation rate in these objects.  

For all star formation scenarios the low Eddington ratio 
sources dominate the AGN contribution to the CIRB at wavelengths $\gtrsim$ 100 $\mu$m.  The AGN contribution to the infrared luminosity 
density is dominated by moderate Eddington ratio AGN at all luminosity levels for all star formation scenarios except 
the AGN evolution star formation model, where the high Eddington ratio sources dominate the ULIRG range.  The star 
formation scenarios that do not include redshift evolution find that low Eddington ratio sources dominate 
the AGN contribution to the local infrared luminosity density.

\subsection{Implications for {\em Herschel} and ALMA}
\label{sub:alma}

The three wide area surveys conducted by {\em Herschel} will yield a large catalog of AGN host galaxies.  Using the planned survey depths and areas as described in Section \ref{sect:intro}, the models discussed here predict the following numbers of AGN hosts to be observed by the {\em Herschel} wide field surveys.  Not including the planned lensing cluster observations, PEP should observe 100-500 AGN hosts, depending on the star formation scenario, at 160 $\mu$m with 5$\sigma$ significance.  The portion of HerMES conducted during the {\em Herschel} science demonstration phase should provide 140-2100 AGN hosts, depending on the star formation scenario, with 5$\sigma$ significance at 250 $\mu$m.  The H-ATLAS survey will yield 250-4000 AGN hosts at 160 $\mu$m and 90-1200 AGN hosts at 350 $\mu$m with 5$\sigma$ significance.  The H-ATLAS observed AGN host counts drop to 30-230 in the 500 $\mu$m band.  This catalog of sources will provide a robust sample to better constrain the star formation properties of AGN hosts.

\citet{Hat10} showed that AGN cannot be differentiated from normal galaxies based only on their FIR colors; therefore, ALMA will only be able to offer supplementary data on AGN which are identified in other wavelength bands.  The wide area surveys conducted by {\em Herschel} will provide many promising candidate targets for ALMA.  The dotted grey lines in figures \ref{fig:dnds350} and \ref{fig:dnds500} show the sensitivity limit of the 350 and 450 $\mu$m ALMA bands, respectively, as quoted in section \ref{sect:intro}.  Based on this projected sensitivity and the models presented here, the areal density of AGN hosts available to ALMA at 450 $\mu$m will be 300-1500 deg$^{-2}$, depending on the star formation scenario, for an integration time of only 60 seconds.  Therefore, using ALMA to conduct follow up observations on X-ray selected AGN should be an efficient way to study star formation in AGN host galaxies.  As discussed in section \ref{sub:sf}, the differential number counts for different star formation scenarios seem to peak at different flux levels.  These peak fluxes become more differentiated at higher wavelengths.  By providing deeper submillimeter observations of {\em Herschel} sources and X-ray selected AGN, ALMA should be able to determine the evolution of star formation in AGN hosts.  This will allow the determination of the wavelength where the AGN contribution to the CIRB peaks, and therefore whether the star formation in AGN hosts evolves differently than in normal galaxies. Furthermore, since ALMA will be taking spectra, measurements such as gas velocity, abundances, and temperature can be made.  This will allow ALMA to not only determine the evolution of star formation in AGN hosts, but also to probe the physical structure of AGN host galaxy star formation.

\subsection{Evolution of AGN}
\label{sub:agnevol}
 
From deep X-ray surveys it is known that high luminosity quasars and moderate luminosity AGN evolve differently 
with respect to redshift \citep[e.g.,][]{U03}.  This would suggest that quasars and Seyferts are caused by 
processes with different time scales.  Also, it appears that the Seyfert population is well described by the 
unified scheme, but the high luminosity AGN population may not follow the unified model \citep[e.g.,][]{B06, L10}.  
A picture is starting to surface where the most powerful AGN are triggered by major mergers as explored in the 
simulations by \citet{H06} and others, but most AGN are triggered by less violent processes \citep{B06a}.  This picture is 
taken into account here by using the composite model of \citet{DB10} to describe the CT AGN fraction as Eddington 
ratio dependent.  The difference in triggering process will also affect the star formation within the host galaxy.  
Mergers will engender not only AGN activity but also bursts of star formation.  Where as the more secular evolution 
experienced by moderate power AGN will induce moderate, constant star formation. 

This AGN evolution scenario was further explored here with the AGN evolution star formation scenario.  In this 
scenario, high Eddington ratio CT AGN hosts were assigned SFR = 175 M$_{\odot}$ yr$^{-1}$.  
The differential counts for this star formation scenario show a large peak $\gtrsim$ 10 mJy.  
However, the high Eddington CT AGN account for $>$ 100$\%$ of the local ULIRG range infrared luminosity density measured by 
\citet{Sey10}.  If the SFR in high Eddington ratio CT AGN hosts is reduced to 100 M$_{\odot}$ yr$^{-1}$, then AGN will 
contribute 80$\%$ of the local ULIRG range infrared luminosity 
density, with a larger contribution to the local LIRG range infrared luminosity density.  The prominent peak in the differential counts 
at $\sim$ 10 mJy persists, despite the reduced SFR in high Eddington ratio CT AGN.  Therefore, a test of the AGN evolution scenario will be if the observed differential counts 
feature a strong bright end peak, possibly with a knee at moderate fluxes.  As a large fraction of the sources contributing to this peak will be CT 
AGN, the AGN samples used for this test will have to be chosen very carefully.  

\section{Summary}
\label{sect:sum}

FIR observations by telescopes like {\em Herschel} and ALMA will provide important insights on many questions about AGN and their hosts.  
Determining the flux level at which the differential AGN and host number counts peak will offer crucial constraints 
to the star formation history of AGN hosts, especially when observing at wavelengths at $\sim$500 $\mu$m.  The predictions 
presented here show that it is likely that the SFR in AGN hosts evolves differently than the SFR in normal galaxies, 
as indicated by the peak wavelength of the AGN contribution to the CIRB being significantly longer than the peak 
wavelength of the CIRB.  Understanding how the SFR evolution in active galaxies differs from quiescent galaxies 
will provide clues on the triggering mechanisms of AGN and how the AGN interacts with the host galaxy.  FIR observations 
will also allow the AGN evolution scenario to be tested by comparing SFRs in bright AGN with different levels of obscuration.  
The relative contributions of AGN with various levels of obscuration to the bright end differential counts will also be an 
important test of the major merger trigger model.  Applying X-ray stacking techniques to bright 350 or 500 $\mu$m sources, 
especially sources with a hot dust component in the SED, will be an efficient way of finding CT AGN.

\acknowledgments
The authors thank A. Poglitsch and I. Valtchanov for information on the PACS and SPIRE filters, P. van Hoof for assistance with Cloudy, and the referee for a useful report that helped to improve this paper.  This work was supported in part by NSF award AST 1008067.


{}


%

\end{document}